\documentclass[twocolumn, tighten]{aastex62}
\graphicspath{{./}{figures/}}

\accepted{2024 February 11}
\submitjournal{ApJ}

\shorttitle{Ce and Nd abundances in the Milky Way Bulge-bar stars}
\shortauthors{Sales-Silva et al.}

\begin{document}

\title{A perspective on the Milky Way Bulge-Bar as seen from the neutron-capture elements Cerium and Neodymium with APOGEE}

\correspondingauthor{J. V. Sales-Silva}
\email{joaovictor@on.br, joaovsaless@gmail.com}

\author{J. V. Sales-Silva}
\affil{Observat\'orio Nacional/MCTI, R. Gen. Jos\'e Cristino, 77,  20921-400, Rio de Janeiro, Brazil}

\author[0000-0001-6476-0576]{K. Cunha}
\affiliation{Steward Observatory, University of Arizona, 933 North Cherry Avenue, Tucson, AZ 85721-0065, USA}
\affil{Observat\'orio Nacional/MCTI, R. Gen. Jos\'e Cristino, 77,  20921-400, Rio de Janeiro, Brazil}

\author[0000-0002-0134-2024]{V. V. Smith}
\affiliation{NOIRLab, Tucson AZ 85719}

\author{S. Daflon}
\affiliation{Observat\'orio Nacional/MCTI, R. Gen. Jos\'e Cristino, 77,  20921-400, Rio de Janeiro, Brazil}

\author{D. Souto}
\affiliation{Departamento de F\'isica, Universidade Federal de Sergipe, Av. Marechal Rondon, S/N, 49000-000 S\~ao Crist\'ov\~ao, SE, Brazil}

\author[0000-0002-0151-5212]{R. Guerço}
\affiliation{Observat\'orio Nacional/MCTI, R. Gen. Jos\'e Cristino, 77,  20921-400, Rio de Janeiro, Brazil}

\author{A. Queiroz}
\affil{Instituto de Astrof\'isica de Canarias, E-38200 La Laguna, Tenerife, Spain}

\author{C. Chiappini}
\affil{Leibniz-Institut f{\"u}r Astrophysik Potsdam (AIP), An der Sternwarte 16, 14482 Potsdam, Germany}

\author{C. R. Hayes}
\affiliation{NRC Herzberg Astronomy and Astrophysics Research Centre, 5071 West Saanich Road, Victoria, B.C., Canada, V9E 2E7}

\author{T. Masseron}
\affil{Instituto de Astrofísica de Canarias (IAC), E-38205 La Laguna, Tenerife, Spain}
\affil{Universidad de La Laguna (ULL), Departamento de Astrofísica, 38206 La Laguna, Tenerife, Spain}

\author{Sten Hasselquist}
\affil{Space Telescope Science Institute, Baltimore, MD, USA}

\author{D. Horta}
\affil{Center for Computational Astrophysics, Flatiron Institute,162 Fifth Avenue, New York, NY 10010, USA}

\author{N. Prantzos}
\affil{Institute d'Astrophysique de Paris, UMR 7095 CNRS \& Sorbonne Universit\'e, 98 bis Blvd Arago, Paris  75014  France}

\author{M. Zoccali}
\affil{Instituto de Astrofísica, Pontificia Universidad Católica de Chile, Vicunã Mackenna 4860, Macul, Casilla 306, Santiago 22, Chile}
\affil{Millenium Institute of Astrophysics (MAS), Nuncio Monseñor Sótero Sanz 100, Providencia, Santiago Chile}

\author{C. Allende Prieto}
\affil{Instituto de Astrofísica de Canarias (IAC), E-38205 La Laguna, Tenerife, Spain}
\affil{Universidad de La Laguna (ULL), Departamento de Astrofísica, 38206 La Laguna, Tenerife, Spain}

\author{B. Barbuy}
\affil{Universidade de São Paulo, IAG, Rua do Matão 1226, Cidade Universitária, São Paulo 05508-090, Brazil}

\author{R. Beaton}
\affil{Space Telescope Science Institute, Baltimore, MD, USA}

\author{D. Bizyaev}
\affil{Apache Point Observatory and New Mexico State University, P.O. Box 59, Sunspot, NM 88349-0059, USA}
\affil{Sternberg Astronomical Institute, Moscow State University, Moscow, Russia}

\author{J. G. Fernández-Trincado}
\affil{Instituto de Astronom\'ia, Universidad Cat\'olica del Norte, Av. Angamos 0610, Antofagasta, Chile}

\author{P. M. Frinchaboy}
\affil{Department of Physics \& Astronomy, Texas Christian University, TCU Box 298840, Fort Worth, TX 76129, USA}

\author{J. A. Holtzman}
\affil{New Mexico State University, Las Cruces, NM 88003, USA}

\author{J. A. Johnson}
\affil{The Department of Astronomy and Center of Cosmology and Astro Particle Physics, The Ohio State University, Columbus, OH 43210, USA}

\author[0000-0002-4912-8609]{Henrik J\"onsson}
\affil{Materials Science and Applied Mathematics, Malm\"o University, SE-205 06 Malm\"o, Sweden}

\author{S. R. Majewski}
\affil{Department of Astronomy, University of Virginia, Charlottesville, VA 22904-4325, USA}

\author{D. Minniti}
\affil{Departamento de Fisica, Universidade Federal de Santa Catarina, Trinidade 88040-900, Florianopolis, Brazil}
\affil{Instituto de Astrofísica, Facultad de Ciencias Exactas, Universidad Andres Bello, Avenida Fernández Concha 700, Santiago, Chile}
\affil{Vatican Observatory, I-V00120 Vatican City State, Italy}

\author{D. L. Nidever}
\affil{Department of Physics, Montana State University, P.O. Box 173840, Bozeman, MT 59717-3840}

\author{R. P. Schiavon}
\affil{Astrophysics Research Institute, Liverpool John Moores University, Liverpool, L3 5RF, UK}

\author{M. Schultheis}
\affil{Université Côte d’Azur, Observatoire de la Côte d’Azur, CNRS, Laboratoire Lagrange, Bd de l’Observatoire, CS 34229, 06304 Nice Cedex 4, France}

\author{J. Sobeck}
\affil{Department of Astronomy, University of Washington, Box 351580, Seattle, WA 98195, USA}

\author{G. S. Stringfellow}
\affil{Center for Astrophysics and Space Astronomy, University of Colorado at Boulder, 389 UCB, Boulder, CO 80309-0389, USA}

\author{G. Zasowski}
\affil{Department of Physics and Astronomy, University of Utah, Salt Lake City, UT 84105, USA}



\begin{abstract}

This study probes the chemical abundances of the neutron-capture elements cerium and neodymium in the inner Milky Way from an analysis of a sample of $\sim$2000 stars in the Galactic Bulge/bar spatially contained within $|X_{Gal}|<$5 kpc, $|Y_{Gal}|<$3.5 kpc, and $|Z_{Gal}|<$1 kpc, and spanning metallicities between $-$2.0$\lesssim$[Fe/H]$\lesssim$+0.5.
We classify the sample stars into low- or high-[Mg/Fe] populations and find that, in general, values of [Ce/Fe] and [Nd/Fe] increase as the metallicity decreases for the low- and high-[Mg/Fe] populations. Ce abundances show a more complex variation across the metallicity range of our Bulge-bar sample when compared to Nd, with the r-process dominating the production of neutron-capture elements in the high-[Mg/Fe] population ([Ce/Nd]$<$0.0).  We find a spatial chemical dependence of Ce and Nd abundances for our sample of Bulge-bar stars, with low- and high-[Mg/Fe] populations displaying a distinct abundance distribution. In the region close to the center of the MW, the low-[Mg/Fe] population is dominated by stars with low [Ce/Fe], [Ce/Mg], [Nd/Mg], [Nd/Fe], and [Ce/Nd] ratios. The low [Ce/Nd] ratio indicates a significant contribution in this central region from r-process yields for the low-[Mg/Fe] population. The chemical pattern of the most metal-poor stars in our sample suggests an early chemical enrichment of the Bulge dominated by yields from core-collapse supernovae and r-process astrophysical sites, such as magneto-rotational supernovae. 

\end{abstract}



\section{Introduction}

Unraveling the nature of the Galactic bulge plays a crucial role in understanding the Milky Way's (MW's) formation and evolution, one of the great questions of contemporary astronomy. Historically, bulges are classified as classical or pseudo-bulges (\citealt{Wyse1997}, \citealt{Kormendy2004}), with two main formation scenarios: accretion of small galactic fragments, or in-situ dynamical evolution of the disk, respectively. Models suggest that the outcome of the two scenarios have different properties, as accretion would produce a pressure-supported, spheroidal bulge, while disk dynamical evolution would end up forming a bar, with or without a boxy/peanut. Recent evidences, however, suggest that the picture might be much less bimodal than this. Indeed, several theoretical works revealed that very weak bars, resembling the properties of spheroids, could be produced in situ, starting from disks with large velocity dispersion, much resembling the Milky Way thick disk (e.g., \citealt{DiMatteo_2016}, \citealt{Debattista_2017}, \citealt{Fragkoudi_2018}). On the other hand, \citet{Saha_2016} argue that a pressure-supported spheroid can be spinned up to some net rotation by the formation of a bar. Star-forming disks at z$\approx$2 show large clumps of star formation in the central region of disks, supporting the hypothesis that most (if not all) bulges, regardless of their final shape, have a strong in-situ component (e.g., \citealt{Tadaki_2017}; see also the review by \citealt{Forster-Schreiber_2020}). Along the same lines, by analyzing MW-like galaxies in fully cosmological simulations, \citet{Fragkoudi_2020} concluded that most of them have in-situ bulges.

There is an inherent difficulty in observing the Galactic bulge due to heavy dust extinction coupled with crowded fields. In the last decades, the details of bulge structure have been revealed thanks to observations of large samples of stars in the infrared, which minimizes the dust extinction effect. Through WISE, VVV, and 2MASS observations in the near-infrared, several studies (e.g., \citealt{McWilliam2010}, \citealt{Wegg2013}, \citealt{Ness2016}) contributed to defining a detailed bulge morphology, characterizing it as a bar with a boxy/peanut morphology and an X-shaped structure, forming an asymmetric box, with one side of the box seeming to be larger than the other due to near-far effect.

Unlike most other galaxies, the MW has the advantage that we can resolve its stellar populations, allowing the study of its stars individually to obtain a detailed chemical characterization of the MW disk, halo, and bulge. 
In general, the stellar population of the Galactic bulge is old (\citealt{Zoccali2003}, \citealt{Clarkson2008}, \citealt{Renzini2018}, \citealt{Surot2019}) and presents a broad metallicity range ($-$1.5$<$[Fe/H]$<$0.5; \citealt{Barbuy2018}). The metal-rich bulge component may be composed of some fraction of young/intermediate stars, as shown by \citet[and references therein]{Bensby2017}{} and supported, to a lower extent, by \citet{Haywood2016}, \citet{Schultheis2017}, and \citet{Bernard2018}. Bi-modality in $\alpha$-elements/Fe versus [Fe/H] (meaning the presence of a low- and high--$\alpha$ population) has also been detected in the MW bulge (e.g., \citealt{Rojas-Arriagada2019}, \citealt{Queiroz2020}), even for stars with high probability of being in the boxy-peanut bulge (\citealt{Queiroz2021}). Thanks to infrared photometry and spectroscopy, several studies have unveiled various stellar populations in the direction of the Galactic bulge, such as globular clusters (e.g. \citealt{Borissova2014}, \citealt{Camargo2018}, \citealt{Gran2019}, \citealt{Camargo2019}, \citealt{Obasi2021}), neutron-capture element rich stars (\citealt{Forsberg2023}, \citealt{Mashonkina2023}), N-rich stars (e.g. 
\citealt{Schiavon2017}, \citealt{Fernandez-Trincado2017}, \citealt{Fernandez-Trincado2022}), Al-enhanced stars (\citealt{Fernandez-Trincado2020a}), and possibly ex-situ stars from the Heracles substructure (\citealt{Horta2021}).

A fascinating aspect of the MW bulge is the joint presence of metal-rich and metal-poor stars. Beyond metallicities, these two bulge populations also show distinct distributions in the chemical abundances of other elements and their kinematics (e.g., \citealt{Portail2017}, \citealt{Zoccali2017}, \citealt{Barbuy2018}, \citealt{Queiroz2021}). However, the chemical patterns of elements heavier than iron in these bulge-bar populations are not well defined due to small, fragmented samples with heterogeneous chemical abundance determinations.

The elements beyond the Fe peak (such as Ce and Nd) are mainly produced by two neutron-capture processes, called r- and s-process. 
Since the gravitational wave detection from the neutron star merger GW170817 (\citealt{Abbott2017}) and subsequent spectroscopic observations (\citealt{Watson2019}), neutron star mergers are probably the main producers of r-process elements, as predicted by modern simulations (\citealt{Thielemann2017}). Recently, \cite{Ekanger2023} confirmed that black hole-neutron star mergers also produce a significant amount of r-process elements through nucleosynthesis yields from numerical simulations, as indicated by previous studies as well (e.g. \citealt{Lattimer1976}, \citealt{Freiburghaus1999}, \citealt{Nishimura2016}, \citealt{Kobayashi2023}). The r-process also happens at other astrophysical sites, such as core-collapse supernovae (e.g., \citealt{Wheeler1998}, \citealt{Ekanger2023} ) and magneto-rotational supernovae 
\citep[e.g.,][]{winteler2012,Reichert2021,Reichert2023}. The contribution of core-collapse and magneto-rotational supernovae to the abundance of r-process elements is currently under debate. \cite{Ekanger2023} concluded that magneto-rotational supernovae typically only synthesize up to A $\approx$ 140.
However, thanks to magneto-rotational supernovae, \cite{Kobayashi2023} can reproduce the chemical pattern of the r-process element Eu in the solar neighborhood, with this r-process source producing most of the europium abundance.  

Based on the solar system abundance distribution, the s-process is divided into three components. Massive stars (M $>$ 8 M$_{\odot}$) produce the weak component of the s-process (nickel to strontium, 60 $\lesssim$ A $\lesssim$ 90), using neutrons provided by $^{22}$Ne($\alpha$,n)$^{25}$Mg reaction, and activated in the He-core and C-shell burning \citep{Pignatari2010}. Via this neutron source, more massive asymptotic giant branch (AGB) stars (M $>$ 4 M$_{\odot}$) also overproduce elements from the weak component around Rb (e.g., \citealt{Garcia-Hernandez2006}, \citeyear{Garcia-Hernandez2013}; \citealt{van-Raai2012}, and references therein).
Low-metallicity \citep{Gallino1998} and low- and intermediate-mass AGB stars \citep{Lugaro2003} synthesize through the s-process the elements of strong (A = 204 - 209) and main (A $>$ 90, including cerium and neodymium) components, respectively. The production of the s-process elements in AGB stars occurs in the He-burning layer during thermal pulses, with the $^{13}$C($\alpha$,n)$^{16}$O reaction as the neutron source.

The astrophysical sites of the s-process are better established than those of the r-process, but there is still a need to elucidate some details of the s-process. Currently, for example, we have the Ba (a s-process dominated element) puzzle, where chemical evolution models do not explain the high Ba abundance in young open clusters \citep{Baratella2021,dorazi2022};   also for field stars in MW disk at lower [Fe/H], \citep{Horta2022}. In addition, the current s-process yield from AGB stars is insufficient to explain the chemical pattern of Ba in some dwarf galaxies \citep{Tarumi2021}. The solution to these incompatibilities may come from the improvement of AGB models (implementation of rotation and magnetic field, \citealt{Battino2019}; \citealt{Vescovi2020}) or the addition of other astrophysical sites for the production of s-process elements (like massive rotating stars or spinstars, \citealt{Choplin2017,Prantzos2018}).

The mixed stellar population with a broad metallicity range in the MW bulge allows us to investigate the s-process and r-process nucleosynthesis in different environments. In this context, the bulge may play an important role in resolving the tensions between the models and the measured s-process abundances. However, in general, the studies of the s-process elements in the bulge suffer from low numbers of stars (\citealt{Johnson2012}, \citealt{Bensby2013}, \citealt{VanderSwaelmen2016}, \citealt{Forsberg2019}, \citealt{Lucey2022}, \citealt{Razera2022}, \citealt{Sestito2023}), preventing a clear picture of the s-process chemical pattern. As an exception, \citet{Duong2019} analyzed the Zr, La, Ce, and Nd abundances for 832 stars in the direction of the bulge at latitudes $-$10$^{\circ}$, $-$7.5$^{\circ}$, and $-$5$^{\circ}$. With their sample ranging from [Fe/H]$\approx-$2.0 to [Fe/H]$\approx$0.5, \citet{Duong2019} found flat [X/Fe] trends of La, Nd, and Ce with [Fe/H], attributing this behavior to the s-process contribution from low-mass AGB stars. However, \citet{Lucey2022} found an increase in the [Ba/Fe] ratio with decreasing metallicity using approximately 200 bulge stars. They related this trend to the Ba production by the r-process. In addition, \citet{Razera2022}  derived high [Ce/Fe] ratios for 58 metal-poor bulge stars, indicating an increase of [Ce/Fe] ratio with decreasing [Fe/H]. Clearly, the scenario of the neutron-capture elements in the MW bulge remains inconclusive. 
In this study, we analyze a significant sample of  $\sim$2000 stars in the inner galaxy/bulge with the aim of improving our knowledge of chemical evolution for Ce and Nd within these stellar populations.

We organized this paper as follows: details of the sample and methodology for determining abundances are presented in Section 2. In Section 3, we segregated the different stellar populations in our sample using population catalogs from the literature, [Fe/H], and the [Mg/Fe], [N/Fe], and [Al/Fe] ratios. In Section 4, we investigate the relationship between the Ce and Nd abundances with [Fe/H], and map these abundances in the bulge-bar. Concluding remarks about our results are found in the last section.

\section{Sample and methodology}

The Apache Point Observatory Galactic Evolution Experiment (APOGEE \citealt{Majewski2017}; APOGEE 2 Majewski et al. 2023 in preparation) was one of the experiments carried out as part of the Sloan Digital Sky Surveys (SDSS III and IV).
The SDSS APOGEE observations occur on the 2.5m and 1.0m telescopes at Apache Point Observatory \citep[New Mexico, USA,][]{Gunn2006,Holtzman2010} and the 2.5m telescope at Las Campanas Observatory \citep[La Serena, Chile,][]{Bowen1973} spanning both hemispheres. In particular, the APOGEE survey is a high-resolution (R$\approx$22,500), near-infrared \citep[1.514 to 1.696 microns,][]{Wilson2019} spectroscopic survey which provided in the data release 17 (DR17) results from detailed spectral analyses of approximately 650,000 stars \citep{Abdurrouf2022}. The APOGEE survey sample covers stars from Galactic populations, in particular the bulge, disk, and halo, as well as dwarf satellite galaxies (Large and Small Magellanic Clouds, Saggitarius, and other dwarf spheroidal satellites).

The heavy extinction and crowding of stars towards the Galactic plane and bulge fields make studying these regions difficult. Thanks to APOGEE NIR spectral range, APOGEE can observe the inner region of the Galaxy, which is not reached by large optical spectroscopic surveys. Here, we use this strength of APOGEE to analyze neutron-capture elements in the Galactic bulge.

APOGEE DR17 spectra were reduced following the recipes described in \citet{Nidever2015} with a small modification in the routines for combining the individual visit spectra and calculating radial velocities \citep{Abdurrouf2022}. In the DR17, the stellar parameters and abundances were derived using the APOGEE Stellar Parameters and Chemical Abundance Pipeline \citep[ASPCAP,][]{GarciaPerez2016}, and the APOGEE line list as described in \citet{Smith2021}.
APOGEE DR17 presents some twenty elemental abundances, including the elements Mg, Al, Mn, and Fe, which are discussed here. This paper's Ce and Nd abundances are from the BAWLAS catalog, as discussed below.

\subsection{The BAWLAS catalog}

The BACCHUS Analysis of Weak Lines in APOGEE \citep[BAWLAS,][]{Hayes2022} catalog\footnote{BAWLAS catalog is available on the SDSS DR17 website: \url{https://www.sdss.org/dr17/data_access/value-added-catalogs/}} presents abundance measurements for the heavy elements Ce and Nd, along with Na, P, S, V, Cu, $^{12}$C/$^{13}$C isotopic ratios, and their associated uncertainties, in as many as $\sim$120,000 red giants, depending on the parameter space.
The BAWLAS catalog derivation of the chemical abundances uses calibrated atmospheric parameters from APOGEE DR17 and the line list from \citet{Smith2021} with the fits between observed and synthetic spectra made through the Brussels Automatic Code for Characterizing High Accuracy Spectra \citep[BACCHUS][]{Masseron2016}. BACCHUS uses MARCS model atmospheres \citep[][]{Gustafsson2008} and the radiative transfer code Turbospectrum \citep[][]{AlvarezPlez1998, Plez2012}. 

To get a sample with precise abundances, \citet{Hayes2022} restricted the APOGEE sample to giants with calibrated ASPCAP stellar parameters between 3500 K $<$ T$_{eff}$ $<$ 5000 K and log g $<$ 3.5 and with high signal-to-noise spectra (S/N$>$150).
The BACCHUS code determines the abundances from different methods, such as equivalent width, central line fit, or minimizing $\chi^2$. Diagnostic flags in BACCHUS indicate whether the abundance estimates made by each method are of good quality. In general, \citet{Hayes2022} defined the $\chi^2$ method as the best for most lines. For some very weak lines, where only the central pixels on a line provide a measurable signal, \citet{Hayes2022} used the ``wln'' method that fits only the central pixel of the analyzed line, unlike the $\chi^2$ method in which the entire line is fitted. 
Finally, \citet{Hayes2022} applied line-by-line zero-point abundance calibrations using solar-neighborhood samples, such that these stars yield solar ([X/M] = 0) abundances for each studied line. 

The neutron capture elements Ce and Nd present multiple measurable absorption lines in the H-Band observed by APOGEE. \citet{Cunha2017} analyzed eight measurable Ce II absorption lines in the APOGEE region. 
However, Ce II lines in the H-Band are blended with several atomic, CO, and CN lines and may have small values of equivalent width  \citep{Cunha2017}, hindering the performance of the ASPCAP pipeline to estimate their abundance. Given this scenario, \citet{Hayes2022} re-analyzed Ce in the APOGEE spectra to improve the DR17 estimates through a careful and particular analysis of four Ce II absorption lines (Table \ref{tab_Ce_Nd_lines}). \citet{Hayes2022} measured line-by-line elemental abundances and used quality flags for each line that identify, for example, when a line is highly overlapped or very weak to realize your analysis.

\citet{Hasselquist2016} found ten measurable Nd II lines in the APOGEE spectral region. However, Nd did not have abundance calculated in DR17 due to Nd II lines being weak and blended with absorption lines from other atomic and molecular species. 
Following the same careful procedure used for Ce, \citet{Hayes2022} estimated Nd abundances for the BAWLAS catalog using three Nd II lines (Table \ref{tab_Ce_Nd_lines}). The other Nd II lines are generally very weak and strongly blended, preventing an accurate estimate of abundance. Because some of the Ce and Nd lines overlap with lines from CNO-containing molecules (e.g., CO, CN, or OH), \citet{Hayes2022} redetermined CNO abundances using several lines of these elements, as indicated in Table 5 of their study.
We use the Ce and Nd abundances from the BAWLAS catalog to explore the neutron-capture elements in the Galactic bulge. 

\begin{table}
	\centering
	\caption{Absorption lines used to estimate the Ce and Nd abundances in the BAWLAS catalog.}
	\label{tab_Ce_Nd_lines}
	\begin{tabular}{lccc} 
		\hline
		Element & $\lambda$ (\AA)  & $\chi$ (eV) & $log$ $gf$ \\
		\hline
        Ce II & 15784.8 & 0.318  &  $-$1.510  \\
        Ce II& 16376.5 & 0.122  &  $-$1.965  \\
        Ce II& 16595.2 & 0.122  &  $-$2.186  \\
        Ce II& 16722.5 & 0.470  &  $-$1.830  \\
        Nd II& 15368.1 & 1.264  &  $-$1.550  \\
        Nd II& 16053.6 & 0.745  &  $-$2.200  \\
        Nd II& 16262.0 & 0.986  &  $-$1.990  \\
		\hline\hline
	\end{tabular}
\end{table}

\subsection{The Bulge-bar sample}

\begin{figure}
	\includegraphics[width=\columnwidth]{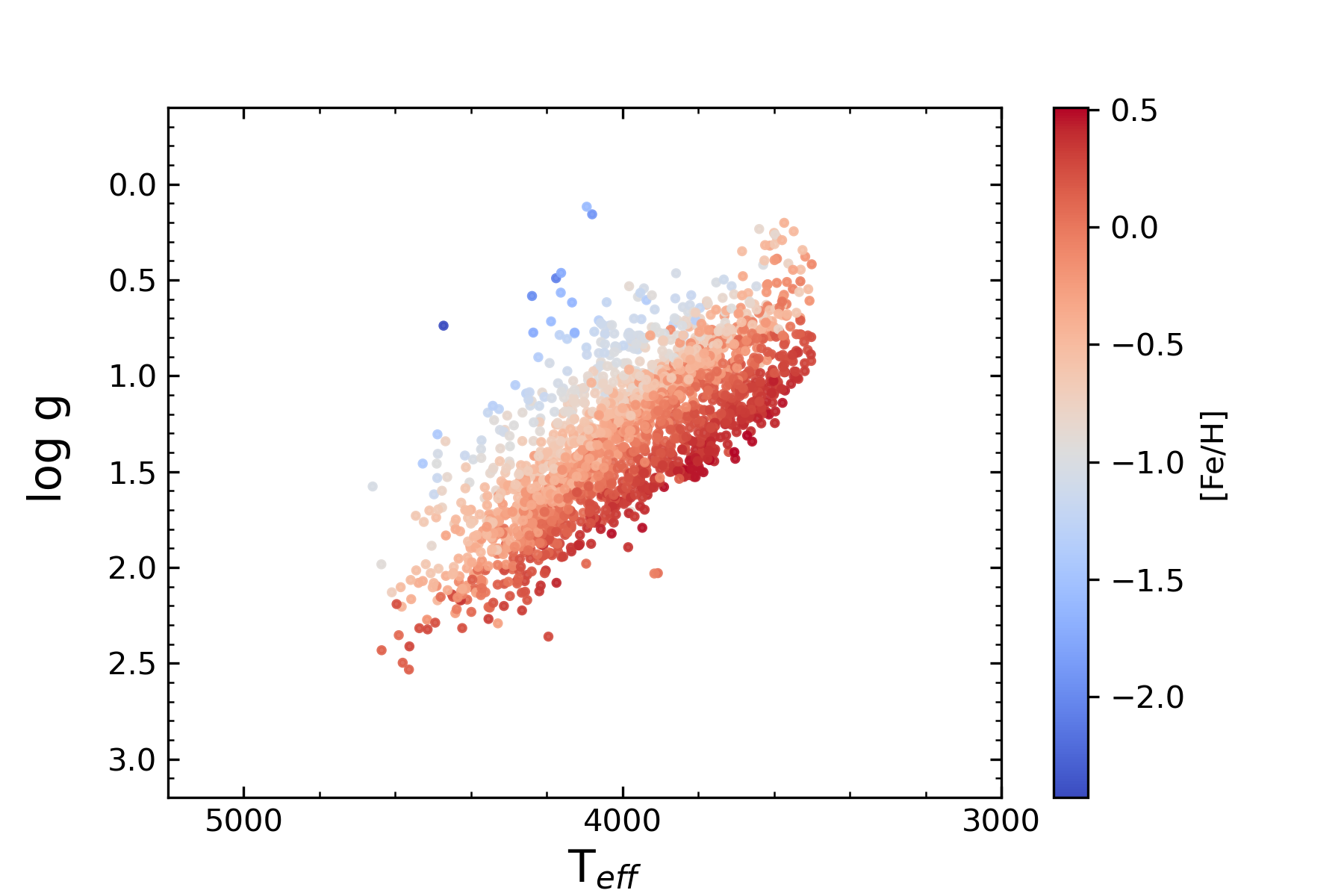}
    \caption{Kiel diagram with colors representing the metallicity for the bulge-bar stars from our sample in the BAWLAS catalog.}
    \label{kiel_diagram}
\end{figure}

\citet{Queiroz2021} investigated the inner region of the Milky Way using data from the APOGEE survey (which included stars in DR16 and DR17 observed until March 2020) combined with photometric surveys, Gaia EDR3 distances and extinctions calculated through the Bayesian StarHorse code \citep{Santiago2016,Queiroz2018}. Their inner galaxy sample contained more than 26,500 stars within the region of $|X_{Gal}|<$5 kpc, $|Y_{Gal}|<$3.5 kpc, and $|Z_{Gal}|<$1 kpc. This sample was cleaned from foreground (disc or halo) stars in \citet{Queiroz2021}  using a reduced-proper-motion (RPM) diagram to separate distinct kinematical populations through very precise proper motions of Gaia EDR3, obtaining a sample of $\sim$8,000 stars considered to be bulge–bar populations. 
In addition, using an orbital diagram, \citet{Queiroz2021} identified different populations (metal-rich thin disk, inner thick disk, stars in bar shape orbits, a more spheroidal pressure supported component, and a counter-rotating stellar population).

Through a cross-match between the BAWLAS catalog and the RPM bulge-bar sample from \citet{Queiroz2021}, we isolated a sample of 2098 stars (henceforward referred to as the bulge-bar-BAWLAS sample), allowing for the study of neutron-capture elements in the bulge.

Figure \ref{kiel_diagram} shows the Kiel diagram using the calibrated atmospheric parameters from APOGEE DR17 of the bulge-bar-BAWLAS stars. The colors in this diagram represent the metallicities of the stars as indicated by the color bar. The blue represents stars in our sample, which are more metal-poor, while red means more metal-rich stars. 

\section{Comparisons with Ce abundances from the literature}

\citet{Hayes2022} found a small systematic difference between the Ce abundances from BAWLAS and DR17 of $\Delta$ ([Ce/H]$_{BAWLAS}$ $-$ [Ce/H]$_{DR17}$) = 0.10$\pm$0.14 for 90,191 stars. When compared to the high-resolution optical study from \citet{Forsberg2019}, there is a very small systematic difference of $\Delta$ ([Ce/H]$_{BAWLAS}$ $-$ [Ce/H]$_{Forsberg}$) = $-$0.04$\pm$0.18 for a sample of 96 stars in common. We note that the Ce abundances for the Milky Way disk from Gaia RVS spectra in \citet{Contursi2023} also do not find systematic differences when compared to \citet{Forsberg2019}.

We compare the Ce abundance results from BAWLAS for the bulge-bar sample with those calibrated values of APOGEE DR17 from the allStar-dr17-synspec\underline{\space}rev1.fits file. The average difference between BAWLAS results and APOGEE DR17 is $\Delta$ ([Ce/Fe]$_{BAWLAS}$ $-$ [Ce/Fe]$_{DR17}$) = 0.08$\pm$0.10.

\citet{Razera2022} redetermined the Ce abundance for 58 metal-poor bulge stars ([Fe/H] $<$ $-$0.8) from \citet{Queiroz2021} RPM sample, using six Ce II lines. To calculate the synthetic spectra, \citet{Razera2022} used the code PFANT \citep{Barbuy2018b} with uncalibrated DR17 atmospheric parameters as input, MARCS atmospheric models, and atomic parameters for Ce II lines from \citet{Cunha2017}. The bulge-bar-BAWLAS sample has 19 stars in common with the \citet{Razera2022} sample, with an average difference of $\Delta$ ([Ce/Fe]$_{Razera}$ $-$ [Ce/Fe]$_{BAWLAS}$) = 0.16$\pm$0.13. We did not find bulge-bar-bawlas stars with estimated Nd abundances in the literature.

\section{A chemical perspective on the bulge-bar stellar populations}

\begin{figure}
	\includegraphics[width=\columnwidth]{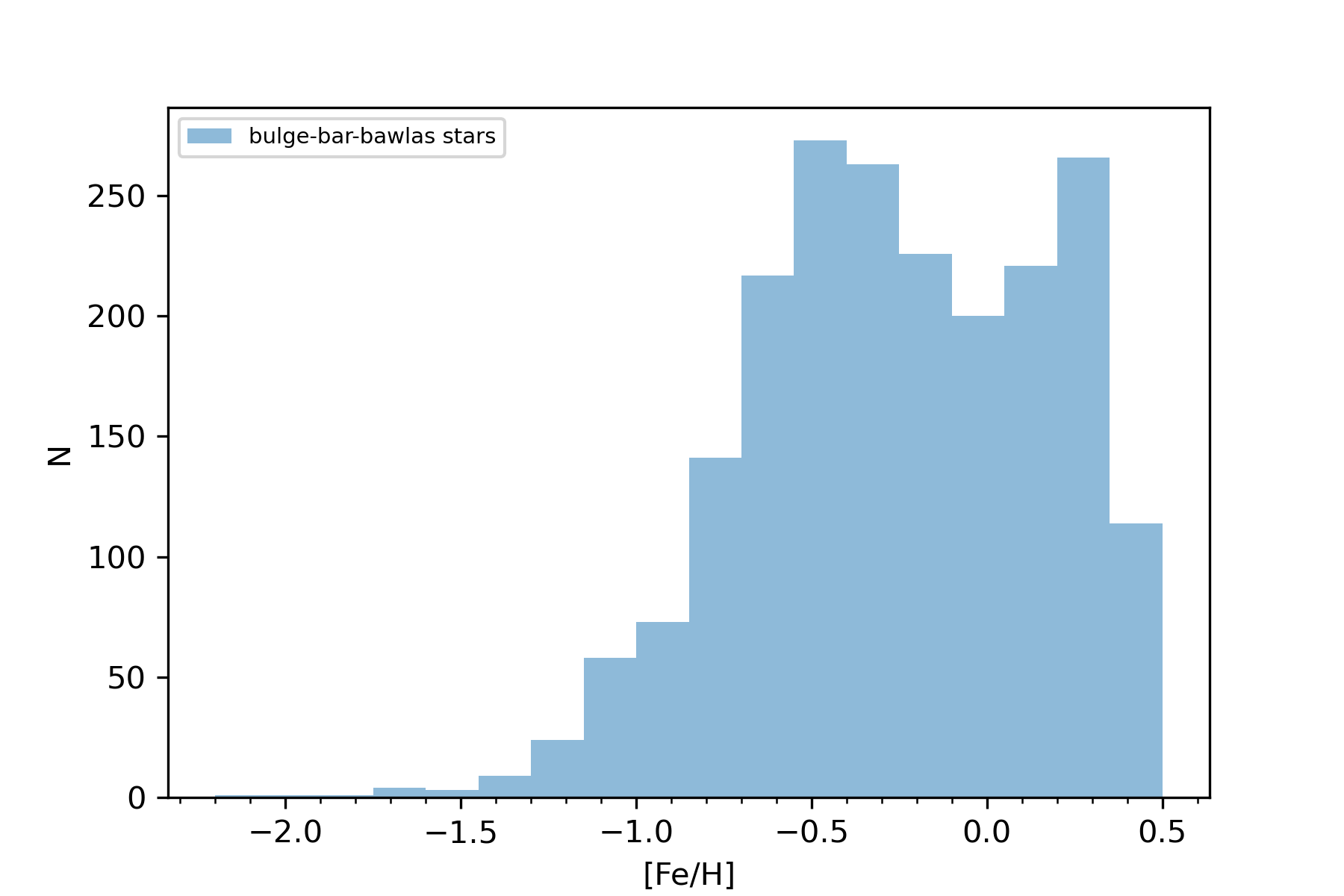}
    \caption{Metallicity distribution for the bulge-bar-BAWLAS sample in this study.}
    \label{Fe_hist}
\end{figure}

The multi-peaked metallicity distribution function (MDF) in the bulge (e.g. \citealt{Zoccali2008}; \citeyear{Zoccali2017}; \citealt{Ness2013}; \citealt{Rojas-Arriagada2017}; \citealt{Johnson2022}) indicates the presence of multiple populations in this Galactic component. The number and location of peaks in the bulge MDF vary across studies, depending on the sample analyzed (see Table 2 in \citealt{Barbuy2018} ).
In Figure \ref{Fe_hist}, we show the metallicity distribution for the bulge-bar-BAWLAS sample. This distribution displays two peaks, a more metal-rich peak around [Fe/H]$\approx$0.3, with a metal-poor peak being wider and centered around $-$0.5, similar to the bulge-bar MDF found in \citealt{Queiroz2021} for a sample of 8,000 bulge-bar stars. 

\begin{figure*}
\centering
	\includegraphics[width=15cm]{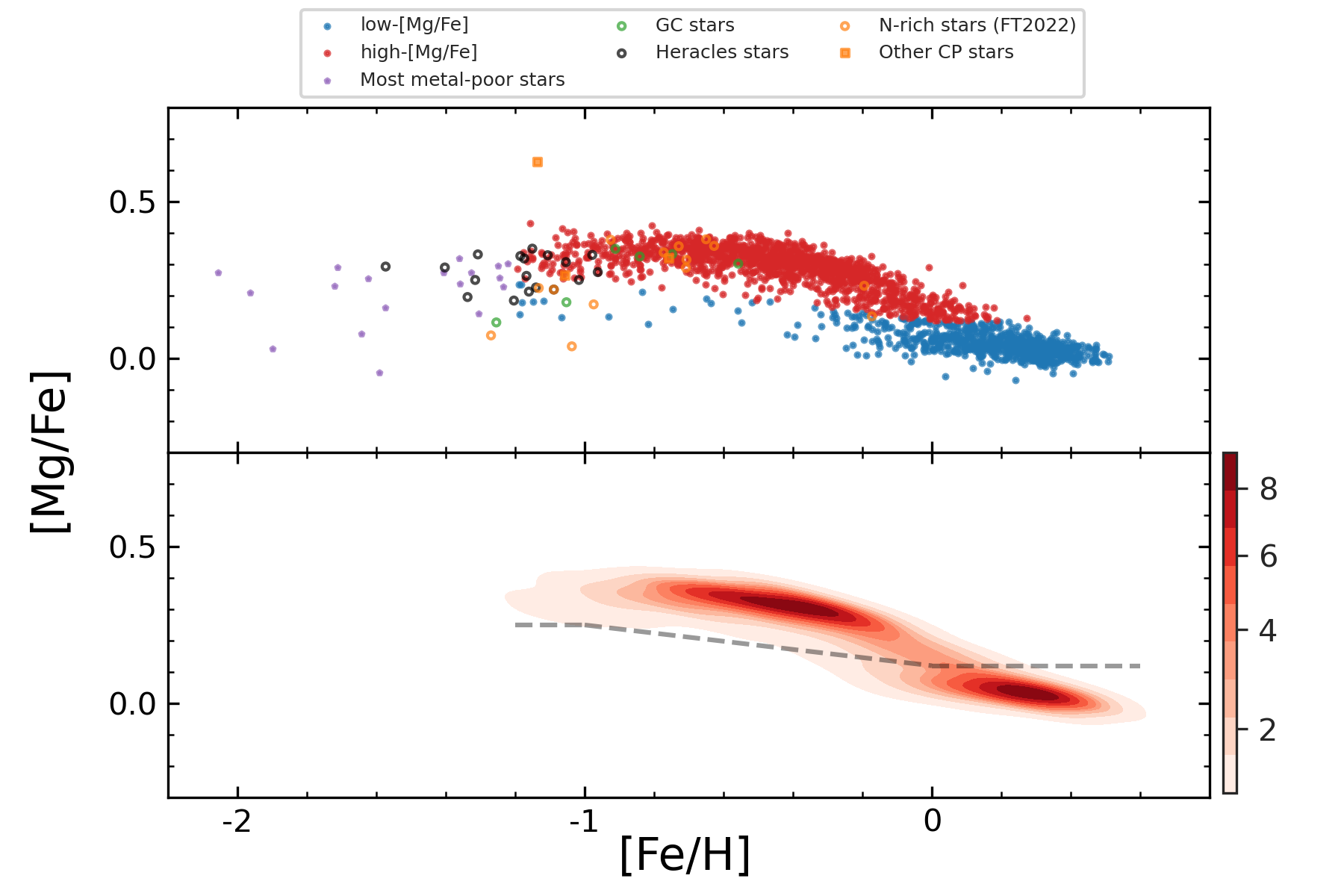}
    \caption{[Mg/Fe]-[Fe/H] plane. The top panel shows the bulge-bar populations present in the BAWLAS catalog, with the legend box above the panel indicating the meaning of the different symbols, read section 4 for classification details. The bottom panel shows the distribution, color-coded according to the probability density function. The dashed line separates the low- and high-[Mg/Fe] populations, as described in the text.}
    \label{MgFe_FeH}
\end{figure*}

We used the calibrated Mg abundances from APOGEE DR17 to investigate the $\alpha$-element abundance patterns in the bulge-bar-BAWLAS sample. 
Figure \ref{MgFe_FeH} illustrates that overall the [Mg/Fe] ratios are different in metal-poor and metal-rich stars in our sample.
The density distribution of the bulge-bar-BAWLAS sample shown in the bottom panel of Figure \ref{MgFe_FeH} extends to around $-$1.2 in [Fe/H] and clearly shows two denser regions, the metal-poor region centered at [Mg/Fe]$\approx$0.3 and the metal-rich one at [Mg/Fe]$\approx$0.0. We used the division line 
in \citet{Weinberg2022} and the density plot shown in the bottom panel of Figure \ref{MgFe_FeH} to separate our sample into high- and low-[Mg/Fe] populations. This division line (shown as dashed line) is described as [Mg/Fe]=0.12 for [Fe/H]$\geq$0, [Mg/Fe]=0.12$-$0.13[Fe/H] for $-$1$<$[Fe/H]$<$0 and [Mg/Fe]=0.25 for $-$1.2$\leq$[Fe/H]$\leq-$1.0. 

In the top panel of Figure \ref{MgFe_FeH}, we show the stars classified as high- and low-[Mg/Fe] populations represented as red and blue circles, respectively. The high [Mg/Fe] population is essentially metal-poor, while the low [Mg/Fe] population is overall metal-rich. 
We consider stars with [Fe/H]$<-$1.2 as a separate population beyond the lower metallicity limit of the density distribution. 
We note that the two over-density features at low [Fe/H] - high [$\alpha$/Fe] and high [Fe/H] - low [$\alpha$/Fe] in Fig. \ref{MgFe_FeH} are also found in the recent theoretical study of \cite{Prantzos_2023} at roughly the same metallicity ranges (see their Fig. 15 for [S/Fe] behavior in the inner disk). In the framework of that model, the inner disk is little affected by radial migration, and the over-density of stars at super-solar metallicities ([Fe/H]$\sim$ +0.2 to +0.5) is attributed to the late saturation of metallicity in that region. In contrast, there is no such over-density at super-solar metallicities in the solar vicinity, and the observed stars in that metallicity range are attributed to radial migration.

The bulge may also be composed of stellar members of Bulge Globular Clusters (\citealt{Bica2016}; \citealt{Geisler2021}; \citealt{Schiavon2017b}), dissipated Globular Clusters (\citealt{Shapiro2010}; \citealt{Kruijssen2015}; \citealt{Bournaud2016};  \citealt{Schiavon2017}), and extra-Galactic stars coming from mergers of the Milky Way with other galaxies, such as Heracles \citep{Horta2021}. 
We also investigated the presence of stars belonging to these populations in our sample.
\citet{VasilievBaumgardt2021} cataloged the probable member stars of 170 globular clusters using data from Gaia Early Data Release 3 (EDR3). We cross-matched the probable members of GCs from \citet{VasilievBaumgardt2021} against our bulge-bar-BAWLAS sample to examine whether the sample contains stars that are probable GC members. We found six such stars in our sample, 2M17333292-3323168 from Liller 1, 2M17342693-3904060 from NGC 6380, 2M18032356-3001588 and 2M18033819-3000515 from NGC 6522, 2M18133324-3150194 from NGC 6569 and 2M17361421-3834371 from Pismis 26; these are shown as green open circles in the top panel of Figure \ref{MgFe_FeH}.

N-rich stars are another population also present in the inner Galaxy (e.g., \citealt{Fernandez-Trincado2016}, \citeyear{Fernandez-Trincado2017}, \citeyear{Fernandez-Trincado2019b}; \citealt{Schiavon2017}; \citealt{Kisku2021}). As indicated by chemical and kinematical similarities between N-rich stars and GC stars (\citealt{Carollo2013}; \citealt{Tang2020}; \citealt{Yu2021}), the nature of N-enhanced stars is probably linked to stripped GCs caused by the interaction of the Milky Way with other galaxies, such as Sagittarius. 
Another possible scenario for forming an N-rich star is mass transfer in a binary, where the N enrichment would result from mass transfer from the evolved companion star, as suggested by \citet{Fernandez-Trincado2019a}.
\citet[FT2022]{Fernandez-Trincado2022} provide a catalog of 412 N-rich stars in the APOGEE survey, fourteen of which are in our sample. 
The orange open circles in the top panel of Figure \ref{MgFe_FeH} represent the N-rich stars from FT2022.

We also analyze the N and Al abundances of our sample to investigate whether other stars, beyond the chemically peculiar stars from FT2022, present high N or Al abundances. Like N, some second-generation stars from GCs have an overabundance of Al compared to Fe (e.g., \citealt{Fernandez-Trincado2019b}, \citealt{Masseron2019}, \citealt{Meszaros2020}). In this analysis, we found two Al- and N-rich stars (2M18032060-3004281 with [Fe/H]=$-$0.76, [Al/Fe]=0.60, and [N/Fe]=1.48; and 2M18091354-2810087 with [Fe/H]=$-$1.14, [Al/Fe]=0.91, and [N/Fe] = 0.81), and one N-rich star (2M17482754-2301526, with [N/Fe] = 1.07 and [Fe/H]=$-$1.06) using the Al calibrated abundances from DR17/APOGEE and N abundances from BAWLAS catalog. These stars can be classified as N-rich stars according to the definition of FT2022 shown in their Figure 1. 
We represent 2M17482754-2301526, 2M18032060-3004281, and 2M18091354-2810087 as orange squares (labeling them as other chemically peculiar (CP) stars) in the top panel of Figure \ref{MgFe_FeH}. 

\citet{Horta2021} discovered a substructure (called Heracles) located in the central region of the Galaxy, probably linked to an accretion event. They defined the stellar population of Heracles as metal-poor, discernible in the [Mg/Mn]-[Al/Fe] plane, and with eccentric and low-energy orbits. \citet{Myeong2022} tried to connect Heracles with another recently discovered population, Aurora, an in-situ stellar component found by \citet{BelokurovKravtsov2022}. However, \citet{Horta2023} found chemical differences between Aurora and Heracles. We also investigated whether our sample contains stars from the Heracles substructure, performing a cross-match between \citet{Horta2021} (approximately 300 stars) and the bulge-bar-BAWLAS sample. We found 18 Heracles stars in the bulge-bar-BAWLAS sample. In Figure \ref{MgFe_FeH}, we represent these stars as black open circles. One Heracles star (2M18124455-2719146) is also classified as an N-rich star by FT2022.

The most metal-poor stars in the bulge have metallicities around $-$3.0 (e.g., \citealt{SchlaufmanCasey2014} and \citealt{CaseySchlaufman2015}, \citealt{Reggiani2020}, \citealt{Sestito2023}). Due to the fast chemical enrichment in the bulge, moderately metal-poor stars with [Fe/H] $\gtrsim$ $-$1.5 may be the oldest bulge stars (\citealt{Chiappini2011}, \citealt{Wise2012}, \citealt{Barbuy2018}, \citealt{cescutti2018}). 
We selected stars with [Fe/H]$<-$1.2, the most metal-poor stars in our sample, to investigate the chemical pattern of the neutron-capture elements Ce and Nd in these ancient stars and compare them to the other populations present in our sample. We consider those stars with [Fe/H]$<$ $-$1.2 that are not classified as N-rich stars, nor as probable members of GCs or Heracles, as the most metal-poor population of the bulge-bar-BAWLAS sample (19 stars). These stars are represented by purple circles in Figure \ref{MgFe_FeH}. The chemical characterization of the bulge-bar-BAWLAS stellar populations allows for a more accurate interpretation of the chemical pattern and evolution of the neutron-capture elements Ce and Nd.

\section{Neutron-capture elements in the Galactic Bulge}

To understand the chemical abundance patterns among the heavy elements, one must consider the various paths possible for their nucleosynthesis. 
In the solar system, Ce and Nd are s-process-dominated elements; their s-/r- contributions to the pre-solar composition are estimated, respectively,  as 85/15 \%  for Ce and 62/38 \% for Nd \citep{Prantzos_2020}. 
They belong to the main component of the s-process, and their solar abundances are largely due to He-burning thermal pulses inside the low- and intermediate-mass AGB stars \citep{Lugaro2003,Cristallo2015}. In other environments, such as the central region of the Galaxy, the s-process contribution to the chemical abundance of the heavy elements or even the dominant astrophysical site that produced them may be different from those inferred for the solar system. 
In particular, rotating massive stars (spinstars) may boost the production of s-process elements in their He-burning cores by mixing, from the H-shell, larger amounts of N (leftover from the previous CNO cycle) into the core; the subsequent transformation of $^{14}$N to $^{22}$Ne through two $\alpha$ captures, enhances the production of neutrons by $^{22}$Ne($\alpha$,n)$^{25}$Mg and thus the production of s-process elements, even up to the second heavy-element abundance peak (in non-rotating massive stars only the first peak is reached).  
Using the nucleosynthesis prescriptions of \citet{Frischknecht2016} for rotating massive stars, \citet{Grisoni_2020} studied the chemical evolution of the neutron capture elements in the bulge. \cite{Kobayashi_2020}, on the other hand, modeled the bulge without considering the spinstars as the s-process source, finding the 2$^{\textrm{\tiny {nd}}}$ peak elements (La, Ba, Ce) underproduced at low metallicities [Fe/H]$\lesssim-$1 (see figure 32 of \citealt{Kobayashi_2020}). The yields of heavy elements depend on the rotational velocity of the massive star and the treatment of mixing, mass loss, etc. in the various stellar models.

Adopting a complete grid (in mass, metallicity, and massive-star rotational velocity) of stellar yields from \cite{Limongi_2018} and \cite{Cristallo2015}, as well as a 1-zone model for the chemical evolution of the solar vicinity, \cite{Prantzos2018} found that spinstars may dominate the production of Ce up to a metallicity of [Fe/H]$\sim-$0.6, where AGB stars become the dominant source for more metal-rich stars, see their figure 16. The situation is similar for Nd, except that the r-process contribution is substantial, and even dominant at such low metallicities. 
\citet{Razera2022} also note that the s-process enrichment present in very metal-poor bulge stars may have a significant contribution from massive rotating stars, based on yields from \citet{Frischknecht2016}.

Despite some mismatches with models (e.g. \citealt{Baratella2021,dorazi2022}), the evolution of s-process elements in the MW disk is relatively well established as we have representative samples of open clusters and field stars with good ages, distances, and estimated abundances (e.g., \citealt{Sales-Silva2022,Ratcliffe2023,casali2023}). Overall, there is a growth of the [s-process/Fe] ratios with decreasing age for the disk populations \citep[e.g.,][]{Maiorca2011,Spina2018,Sales-Silva2022,Horta2022} with this relation presenting a metallicity dependence \citep[e.g.,][]{Delgado-Mena2019,Casamiquela2021,Sales-Silva2022}. 
This dependence is due to a complex interplay between the metallicity dependence of the s-process yields (which is different for the various s-process elements) and the delay in enriching the interstellar medium from low-mass AGB stars. 

In this section, we use the Ce and Nd abundances obtained for the multiple stellar populations present in the bulge-bar-BAWLAS sample to investigate the behavior of these elements with metallicity and position of the stars.

\subsection{Ce and Nd abundances versus [Fe/H]}

\begin{figure*}
\centering
	\includegraphics[width=19cm]{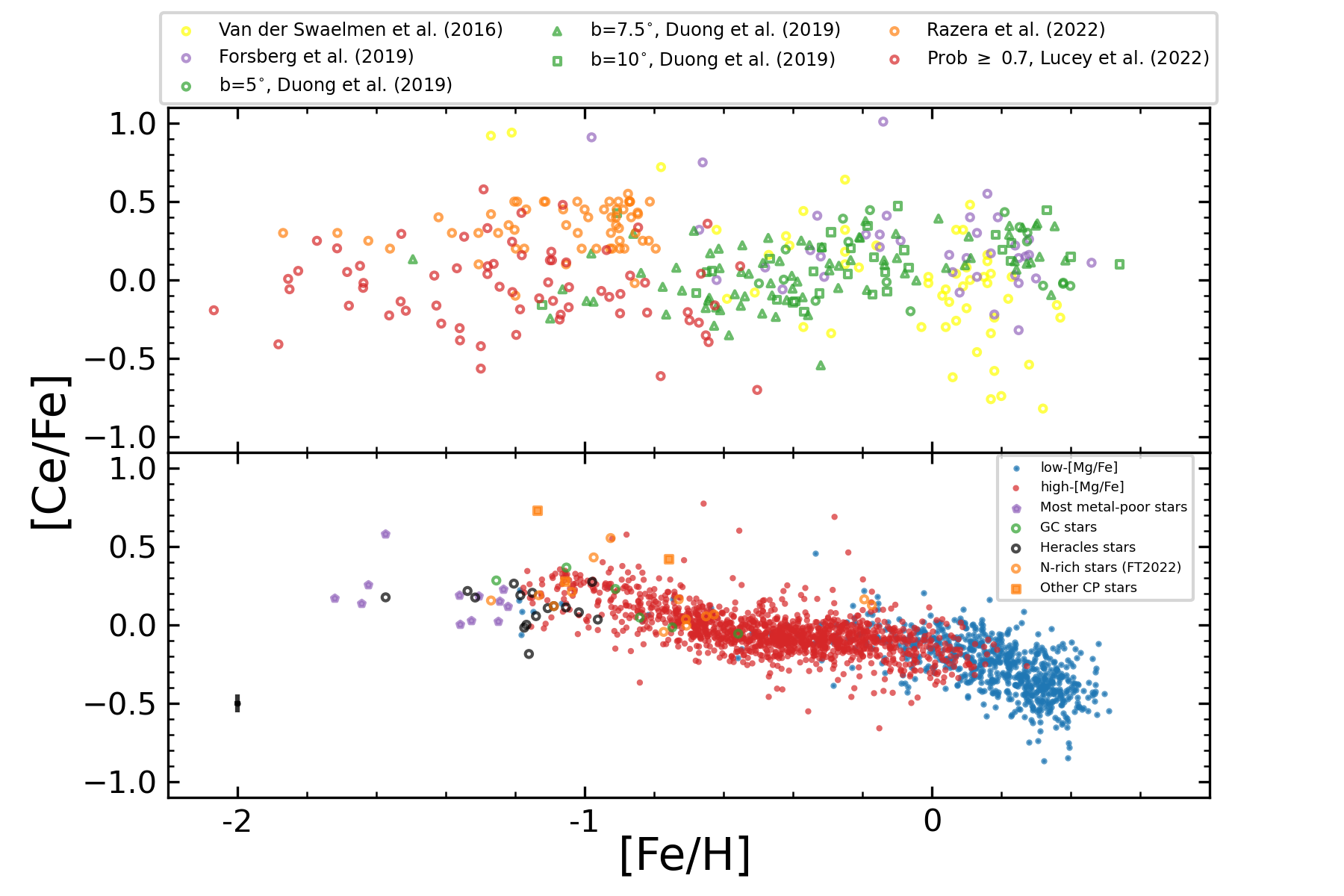}
    \caption{[Ce/Fe]-[Fe/H] plane. The top panel shows results from literature for bulge stellar population: green symbols represent the bulge giant stars observed at latitudes b=$-$10$^{\circ}$ (square), $-$7.5$^{\circ}$ (triangle), and $-$5$^{\circ}$ (circle) from \citet{Duong2019}; yellow, purple, orange, and red circles correspond to bulge giant stars from \citet{VanderSwaelmen2016,Forsberg2019,Razera2022,Lucey2022}, respectively. We plot only stars of the \citet{Lucey2022} sample with a probability of belonging to the bulge $\geq$ 0.7. The bottom panel presents the [Ce/Fe] results for the bulge-bar-BAWLAS population, with the legend box inside this bottom panel indicating the meaning of the different symbols. A typical error bar is shown at the bottom left.
    }
    \label{Ce_Fe_vs_Fe_H}
\end{figure*}

\begin{figure*}
\centering
	\includegraphics[width=19cm]{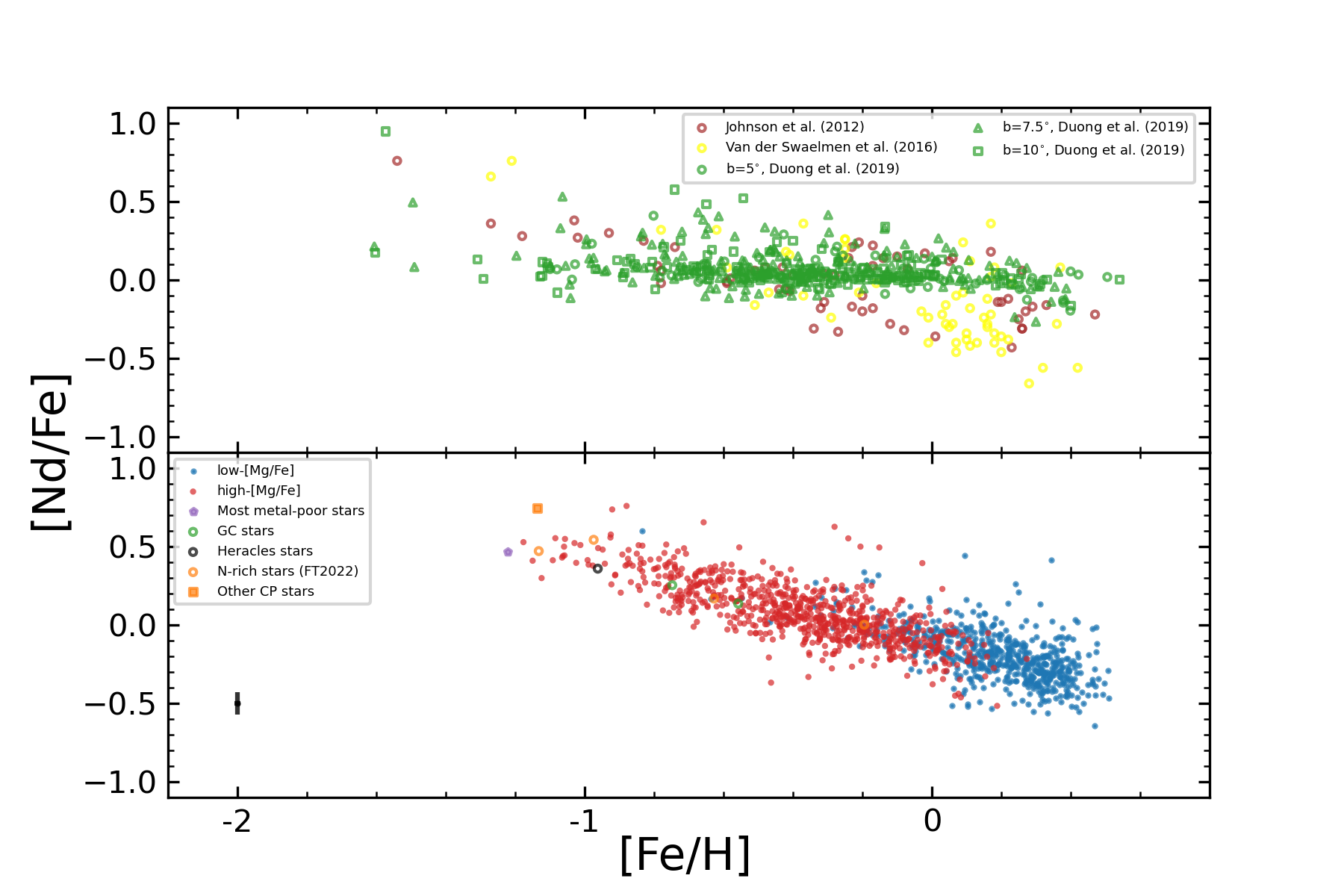}
    \caption{[Nd/Fe]-[Fe/H] plane. The top panel shows bulge-bar results from the literature, while the bottom panel displays [Ce/Nd] ratios of the bulge-bar stars from the BAWLAS catalog. The symbols present the same meaning as Figure \ref{Ce_Fe_vs_Fe_H}, with the addition of brown circles in the top panel representing bulge stars from \citet{Johnson2012}. A typical error bar is shown in the bottom panel.
    }
    \label{Nd_Fe_vs_Fe_H}
\end{figure*}

\begin{figure*}
\centering
	\includegraphics[width=19cm]{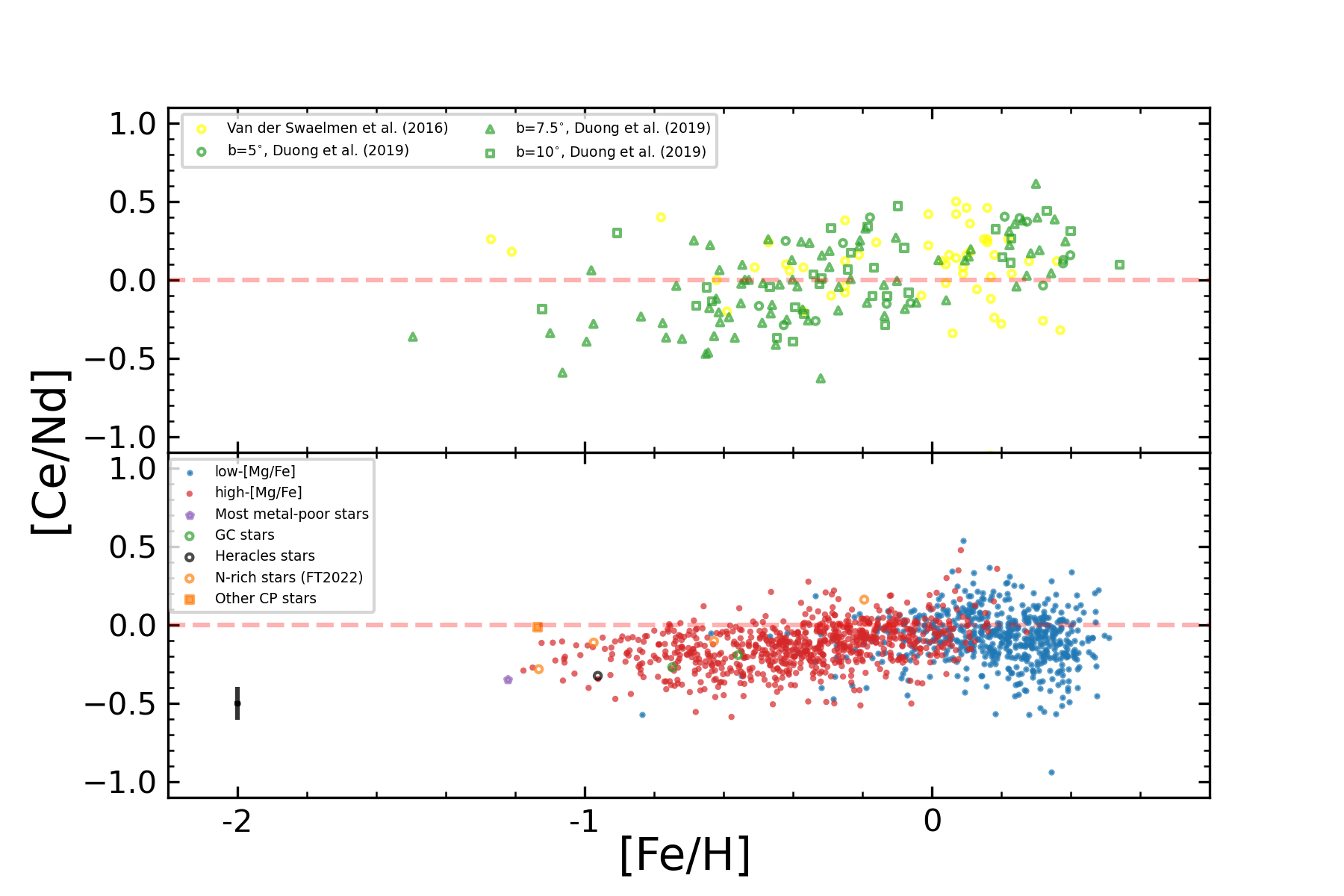}
    \caption{[Ce/Nd]-[Fe/H] plane. The top panel shows bulge-bar results from literature, while the bottom panel displays [Ce/Nd] ratios of the bulge-bar stars from the BAWLAS catalog. The symbols present the same meaning as Figure \ref{Ce_Fe_vs_Fe_H}. A typical error bar is shown in the bottom panel. 
    }
    \label{Ce_Nd_vs_Fe_H}
\end{figure*}

In the bottom panels of Figures \ref{Ce_Fe_vs_Fe_H}, \ref{Nd_Fe_vs_Fe_H} and \ref{Ce_Nd_vs_Fe_H}, we present [Ce/Fe], [Nd/Fe], and [Ce/Nd] ratios versus [Fe/H], respectively, for the stellar populations in the bulge-bar-BAWLAS sample. The symbols in these bottom panels represent the low- (blue circles) and high-[Mg/Fe] population (red circles), the most metal-poor stars (purple pentagons), the GC stars (open green circles), the Heracles stars (open black circles), N-rich stars (from FT2022; open orange circles) and other chemically peculiar stars (orange squares).

The bulge results of [Ce/Fe], [Nd/Fe], and [Ce/Nd] ratios from the literature are shown in the top panels of Figures \ref{Ce_Fe_vs_Fe_H}, \ref{Nd_Fe_vs_Fe_H}, and \ref{Ce_Nd_vs_Fe_H}, respectively. Green symbols represent the bulge giant stars observed at latitudes b=$-$10$^{\circ}$ (square), $-$7.5$^{\circ}$ (triangle), and $-$5$^{\circ}$ (circle) from 
\citet{Duong2019}. While brown, yellow, purple, orange, and red circles correspond to bulge giant stars from \citet{Johnson2012, VanderSwaelmen2016,Forsberg2019,Razera2022,Lucey2022}, respectively. For the \citet{Lucey2022} sample, we selected only stars with a probability of belonging to the bulge $\geq$ 0.7.

\subsubsection{[Ce/Fe]-[Fe/H] plane}

The literature results for the [Ce/Fe] ratio in the bulge shown in the top panel of Figure \ref{Ce_Fe_vs_Fe_H} extend over a large metallicity range between $-$2.1$\lesssim$[Fe/H]$\lesssim$0.5. The combined view of these results from different chemical studies 
shows a large dispersion in the abundances, which may be due, in part, to abundance uncertainties and systematics in the abundance determinations that were derived using different methods.

The top and bottom panels of Figure \ref{Ce_Fe_vs_Fe_H} reveal a different and tighter overall behavior for [Ce/Fe] versus [Fe/H] for the bulge-bar-BAWLAS sample with less scatter at any given metallicity when compared to literature studies. \citet{Forsberg2019} and \citet{Lucey2022} derived a flat trend of [Ce/Fe] vs [Fe/H] throughout the metallicity range of their bulge sample. Differently, \citet{Duong2019} found [Ce/Fe] solar for [Fe/H]$<$0 and high [Ce/Fe] ratios for bulge stars with [Fe/H]$>$0. 
\citet{VanderSwaelmen2016}  found high [Ce/Fe] ratios in the metal-poor regime and a decrease in [Ce/Fe] as the metallicity increases. 
As previously mentioned, \citet{Razera2022} focused on the analysis of low metallicity bulge APOGEE data but, unlike BAWLAS, used the ASPCAP uncalibrated parameters to derive Ce abundances, finding Ce enrichment with a mean of [Ce/Fe]$\approx$0.4 in the metal-poor bulge stars ([Fe/H]$<-$0.8). Their results have a mean offset of 0.16 compared to ours, which is not surprising, given the different stellar parameters scale.

For the bulge-bar-BAWLAS sample, the relationship between [Ce/Fe] and metallicity shows smooth variations in the trend over the metallicity range (see bottom panel of Figure \ref{Ce_Fe_vs_Fe_H}). In the metal-rich regime ([Fe/H]$>$0), the low-[Mg/Fe] bulge population displays an increase in the [Ce/Fe] ratio as the metallicity decreases, finding a flat trend with [Ce/Fe]$\approx-$0.10 for $-$0.3$\lesssim$[Fe/H]$\lesssim$0. For [Fe/H]$<-$0.3, the low-[Mg/Fe] population in the bulge-bar-BAWLAS sample has few stars, but in general, these stars follow the trend shown by the high-[Mg/Fe] population.

The high-[Mg/Fe] stars follow an approximately constant trend ([Ce/Fe]$\approx-$0.10) for $-$0.6$\lesssim$[Fe/H]$\lesssim$0.2, followed by an increase in the [Ce/Fe] ratio with a decrease in metallicity for $-$1.0$\lesssim$[Fe/H]$\lesssim-$0.6. Finally, the high-[Mg/Fe] population also shows a ``knee'' feature in [Fe/H]$\approx-$1.0, displaying a flat trend between $-$1.2$\lesssim$[Fe/H]$\lesssim-$1.0 with [Ce/Fe]$\approx$0.2. The most metal-poor stars in the bulge-bar-BAWLAS sample extend this constant trend of [Ce/Fe] for [Fe/H]$<-$1.2, differently to that presented by the Milky Way Satellites, like Sagittarius Dwarf Galaxy, which shows a decrease in the [Ce/Fe] ratio in these metallicities \citep{Hasselquist_2021}.
In the metallicity range where the low- and high-[Mg/Fe] populations overlap, these populations are located in the same region of the [Ce/Fe]-[Fe/H] plane, sharing a similar trend.

We note that the [Ce/Fe] ratio declines smoothly with metallicity, but this decline makes a ``pause'' in the region of [Fe/H]$\sim-$0.5 to 0. Although the statistical significance of that feature is small, from the theoretical point of view, a more or less broad bump is expected in that metallicity region. Indeed, the efficiency of the s-process in low-mass stars is known to increase with metallicity and go through a maximum at a value that depends on the mass number of the isotope: the production of heavier isotopes reaches that maximum at lower metallicity values than lighter ones \citep[see, e.g., Fig. 1 in][]{Travaglio_2004}. For that reason, the evolution of the [s-/Fe] ratio of elements of the second s-peak, like Ba or Ce, is expected to display a small ``bump'' at slightly subsolar metallicities, as found in Fig. 5 of \citet{Travaglio_2004}, Fig. 2 of \cite{Grisoni_2020}, or in \citet{Prantzos2018} for 1-zone models (their Fig. 16) and \citet{Prantzos_2023} (their Fig. 14) which adopt different sets of yields (see discussion in first paragraph Sec. 5).

\subsubsection{[Nd/Fe]-[Fe/H] plane}

The top panel of Figure \ref{Nd_Fe_vs_Fe_H} shows the literature results in the [Nd/Fe] vs [Fe/H] plane for bulge studies from the literature, with different studies indicating overall distinct trends between Nd and metallicity. \citet{Johnson2012} and \citet{VanderSwaelmen2016} found a clear increase in the [Nd/Fe] ratio with decreasing metallicity, being similar to the trend found for the [Eu/Fe] ratio in the bulge. 
In fact, the contributions of s- and r- processes to Nd \citep[62 \% and 38 \%, respectively, according to][]{Prantzos_2020} concern the solar composition, the r- contribution is obviously higher at lower metallicities. 
On the other hand, with a sample of 832 bulge giant stars, \citet{Duong2019} derived an approximately flat trend between [Nd/Fe] and [Fe/H].

In the lower panel of Figure \ref{Nd_Fe_vs_Fe_H}, we show the [Nd/Fe] versus [Fe/H] for the bulge-bar-BAWLAS sample. We find that the [Nd/Fe] ratio decreases as the metallicity increases for the low- and high-[Mg/Fe] populations, presenting a simple linear relationship with metallicity, unlike the [Ce/Fe] ratio, which shows a trend variation in the bulge metallicity range. This different behavior between Ce and Nd is likely linked to the different r-/s-process fractions that contribute to the abundances of these elements, with the Nd trend more closely resembling r-process dominated elements, such as Eu, as shown by \citet{Johnson2012} and \citet{VanderSwaelmen2016}. Only one bulge-bar star with a metallicity lower than $-$1.2 had the [Nd/Fe] ratio estimated in the BAWLAS catalog, finding a high [Nd/Fe] value for this star.

The results in our study indicate positive [Ce/Fe] and [Nd/Fe] for low metallicity bulge stars. As discussed previously, spinstars produce larger amounts of s-process elements (mostly of the 1st peak, but also of the 2nd peak) than non-rotating stars at low-metallicities \citep{Prantzos2018}. \citet{Prantzos2018} and \cite{Grisoni_2020} modeled the chemical evolution of neutron-capture elements, implementing yields from spinstars of \citet{Limongi_2018} and \citet{Frischknecht2016}, respectively. 
\citet{Rizzuti_2019} present a detailed comparison between these two different nucleosynthesis prescriptions from \citet{Limongi_2018} and \citet{Frischknecht2016}, indicating that both show better agreement with the neutron capture abundances than in the case when rotation is not included. 
The yields of \citet{Limongi_2018} used in \citet{Prantzos2018} concern rotational velocities up to 300 km/s, being the only grid of yields available for all masses and metallicities and three velocities (0, 150, and 300 km/s). This grid allowed \citet{Prantzos2018} to introduce an Initial Distribution of Rotational Velocities (IDROV) dependent on metallicity and constrained by observations using all elements. 

To summarize, it is difficult at this point to evaluate the exact contribution of spinstars to the abundances of heavy elements at low metallicities. The magneto-rotational supernovae can also contribute to producing neutron-capture elements, such as Ce and Nd, and significantly enrich the bulge in low metallicity, as shown in \citet{Kobayashi_2020} by changing the fraction of these supernovae. The positive [X/Fe] ratios for low-metallicity bulge stars may be a joint consequence of the production of neutron-capture elements by spinstars and an efficient r-process astrophysical site, such as magneto-rotational supernovae. 

Another point to note concerning the bottom panels of Figures \ref{Ce_Fe_vs_Fe_H} and \ref{Nd_Fe_vs_Fe_H} is that, overall, the GC, Heracles, and N-rich stars in the bulge-bar-BAWLAS sample overlap with the high-[Mg/Fe] population in the [Ce/Fe]- and [Nd/Fe]-[Fe/H] planes. In particular, Heracles stars have [Ce/Fe]$>$0 ($<$[Ce/Fe]$>$=0.11$\pm$0.11) in line with the other bulge stars with similar metallicities. Only one star from Heracles (2M17324437-1649444) had a measured Nd abundance in the bulge-bar-BAWLAS sample, with [Nd/Fe]$\approx$0.35. The N-rich, Si-rich star 2M18091354-2810087 \citep{Fernandez-Trincado2020b} also has a high value of [Ce/Fe] ($\approx$0.73) and [Nd/Fe] ($\approx$0.74) compared to bulge stars with the same metallicity, as also found by \citet{Fernandez-Trincado2020b}. 

\subsubsection{The [Ce/Nd]-[Fe/H] plane}

Due to the different contributions from neutron capture processes to Ce and Nd \citep[e.g., 85 \% and 62 \% corresponding to s-process in the solar system, respectively, according to][]{Prantzos_2020}, the abundance ratio [Ce/Nd] in stellar populations indicates which of these processes is dominant in stars. A high [Ce/Nd] ratio indicates a greater proportion of the s-process in the formation of the elements, while the opposite (low [Ce/Nd]) shows a production dominated by the r-process. Here, we investigate the [Ce/Nd] ratio to understand the r-/s-process contribution for the different bulge-bar populations at distinct metallicities. 

Focusing first on the literature [Ce/Nd] ratios shown in the top panel of Figure \ref{Ce_Nd_vs_Fe_H}, overall
\citet{VanderSwaelmen2016} results indicate an approximately constant [Ce/Nd] ratio over the metallicity range of their sample, with most stars showing [Ce/Nd] $>$ 0, which suggests an s-process dominated Bulge population. With a more significant Bulge sample, \citet{Duong2019} found a different trend, in which the most metal-rich stars ([Fe/H]$>$0) show values [Ce/Nd] $>$ 0, while the metal-poor stars exhibit [Ce/Nd]$<$0. Their results show a clear tendency for the [Ce/Nd] ratio to increase with increasing metallicity, which indicates that metal-poor bulge stars have Ce and Nd produced by the r-process.

For the bulge-bar-BAWLAS sample (bottom panel of Figure \ref{Ce_Nd_vs_Fe_H}), in general, we detect an increase in the [Ce/Nd] ratio with [Fe/H] for the high-[Mg/Fe] bulge population, less pronounced but overall in agreement with \citet{Duong2019} up to roughly solar metallicities.
For the low-[Mg/Fe] bulge population, our results show slightly more scatter and possibly a subtle increase followed by a downturn in the [Ce/Nd] ratios from solar to super-solar metallicities, a pattern that is also seen for the Milky Way disk population in Hayes et al. (in preparation).
Our results, in particular for the high-[Mg/Fe] bulge population, indicate an r-process predominance in the Ce and Nd production in the bulge population (overall [Ce/Nd]$<$0), with a small portion of stars, mainly from the low-[Mg/Fe] population having [Ce/Nd] $>$ 0. This scenario indicates that the r-component dominated the neutron capture yields of Ce and Nd at early times in the bulge. The Ce and Nd production observed in metal-poor bulge stars probably comes from events where the r-process occurs: neutron star- and black hole-neutron star mergers \citep{Wanajo2014,Wanajo2021}, core-collapse supernovae \citep{Wanajo2018}, or magneto-rotational supernovae \citep{Nishimura2015}. As pointed out by \citet{Lucey2022}, the source of the r-process enrichment in old, metal-poor bulge stars is unlikely to be the neutron star mergers because these events have a long time-scale 
\citep[$\gtrsim$ 4Gyr,][]{Skuladottir2020}. Magneto-rotational supernovae \citep{Kobayashi2023,winteler2012,Reichert2021,Reichert2023} can produce the heavy elements of the r-process in a shorter time scale than neutron stars and enrich the metal-poor stars of the bulge \citep{cescutti2018}.

Beyond Ce and Nd, the chemical abundances of other heavy elements in bulge-bar stars also indicate a significant contribution of the r-process in low metallicity regimes. The [Eu/Fe] ratio increases as the metallicity decreases for bulge stars \citep{VanderSwaelmen2016,Duong2019}. In addition, \citet{Duong2019} found [La/Eu] $<$ 0 for most bulge stars, with La and Eu being s-process and r-process dominated elements, respectively, indicating a r-process predominance in the abundances of heavy elements in these stars. Recently, \citet{Lucey2022} attributed the Ba enrichment in metal-poor bulge stars to the r-process due to the shorter time scale necessary to enrich these old stars.

\subsubsection{Other Ce and Nd abundance planes}

To compare the contributions of neutron capture processes with core-collapse and Type Ia supernova yields in the bulge population, we investigate Ce and Nd abundances relative to Mg \citep[a tracer of core-collapse supernovae,][]{Woosley-Weaver1995} and Mn \citep[a pristine tracer of supernovae type Ia,][]{iwamoto1999,kobayashi2006}. Iron is produced in supernovae type Ia, like Mn, but presents other relevant production in core-collapse supernovae \citep[e.g.][]{Rodriguez2023}.

\begin{figure*}
\centering
	\includegraphics[width=15cm]{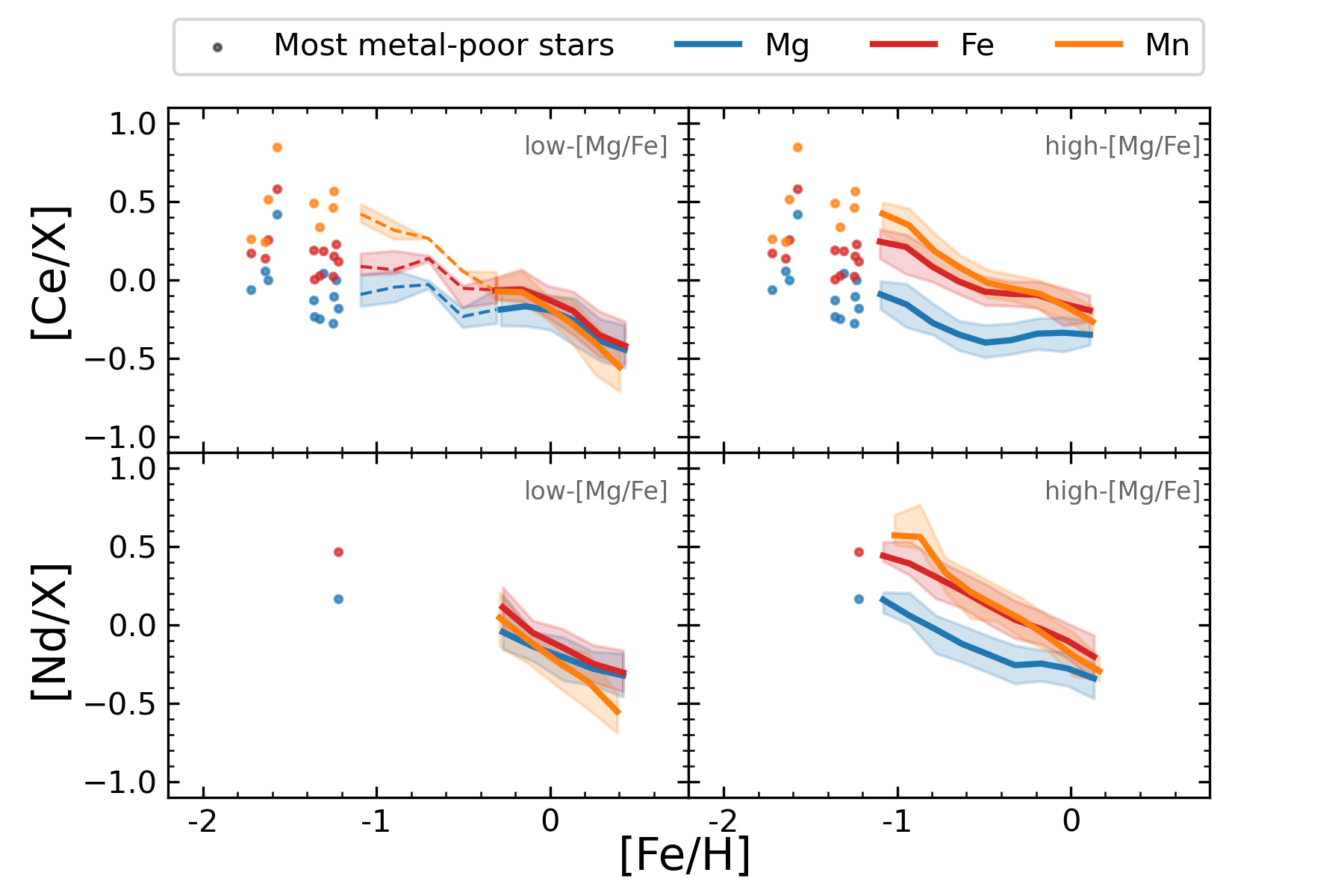}
    \caption{[Ce/X] (top panels) and [Nd/X](bottom panels) as a function of [Fe/H], with X being Mg (blue), Fe (red), and Mn (orange). The lines represent the median trend, and the shaded regions trace the 1$\sigma$ values, with the low- and high-[Mg/Fe] population trends shown in the left and right panels, respectively. Dashed lines in the top left panel represent the metallicity range with few stars. Circles in all panels represent the most metal-poor stars, with colors (blue, red, and orange) indicating the abundance ratio (Mg, Fe, and Mn, respectively).
    }
    \label{Ce_X_vs_Fe_H}
\end{figure*}

In the different panels of Figure \ref{Ce_X_vs_Fe_H}, we show the [Ce/X] and [Nd/X] ratios as a function of [Fe/H] for the most metal-poor stars (filled circles) and the low- and high-[Mg/Fe] populations, with X being Mg, Fe, and Mn. The orange, red, and blue lines represent the median trend of Ce and Nd relative to Mn, Fe, and Mg, respectively, with the shaded regions tracing the standard deviation values. 
Dashed lines represent metallicity ranges with very few stars.

For both the low- and high-[Mg/Fe] populations, overall, the [Ce/X] and [Nd/X] ratios for Mg, Fe, and Mn show an increase as the metallicity decreases. 
The trends of [Ce/Mn] (orange line) and [Ce/Fe] (red line) for the low- and high- [Mg/Fe] populations (shown in the top panels of Figure \ref{Ce_X_vs_Fe_H}) are similar for [Fe/H] $\gtrsim$ $-$0.5 and differ for [Fe/H] $\lesssim$ $-$0.5. Although, for low [Mg/Fe] stars, there is a small number of stars below [Fe/H] $\lesssim$ $-$0.5 (dashed lines). 
The [Ce/Mg] ratio (blue line) for the high-[Mg/Fe] population (top right panel) shows lower values than the [Ce/Fe] and [Ce/Mn] ratios over the metallicity range, as also found (see bottom panels of Figure \ref{Ce_X_vs_Fe_H}) for [Nd/Mg] (blue line) relative to the [Nd/Fe] (red line) and [Nd/Mn] (orange line) ratios. 
For the low-[Mg/Fe] stars (see bottom left panel of Figure \ref{Ce_X_vs_Fe_H}), the curves relative to the [Nd/Mg], [Nd/Fe], and [Nd/Mn] ratios generally overlap in the [Nd/X]-[Fe/H] plane. 
 
The upper right panel of Figure \ref{Ce_X_vs_Fe_H} shows that in the high-[Mg/Fe] stars, the [Ce/Mg] ratio (blue line) displays a flat trend for [Fe/H]$\gtrsim-$0.5 and is inversely proportional to metallicity for [Fe/H]$\lesssim-$0.5.
This change in the trend between [Ce/Mg] and [Fe/H] may result from a delayed s-process source and/or a change in the efficiency of Ce production. The production of s-process elements in AGB stars depends on metallicity \citep{Gallino2006,Cristallo2015,KarakasLugaro2016,Battino2019}, due to AGB yields being related to the number of neutrons provided by the neutron source, probably  $^{13}$C($\alpha$,n)$^{16}$O, per iron-56 seed nucleus. Also, Ce sources through the r-process in neutron star mergers and s-process in low-mass AGBs have gigayear time delays in enriching the interstellar medium. 

The small number of very metal-poor stars in Figure \ref{Ce_X_vs_Fe_H} (plotted as individual filled circles due to small numbers not suitable for medians) are dominated by values of [Ce/Mg]$\leq$0, [Ce/Mn]$>$0.2, and [Ce/Nd]$<$0, suggesting that the early chemical enrichment of the Bulge was dominated by yields from core-collapse supernovae ([Ce/Mg]$\leq$0) and r-process astrophysical sites ([Ce/Mn]$>$0.2, and [Ce/Nd]$<$0.0) which, in this case, are possibly magneto-rotational supernovae due to a shorter timescale compared to neutron star mergers. We estimated the [Nd/Fe] and [Nd/Mg] ratios only for one metal-poor star with [Fe/H] $< -$1.2 from the bulge-bar-BAWLAS sample, 2M18090655-1526144, which has a ratio [Nd/Fe] and [Nd/Mg] equal to 0.47 and 0.16, respectively. (The Mn abundance of 2M18090655-1526144 was not estimated in APOGEE DR17.)

\subsection{Mapping Ce and Nd abundances in the Galactic bulge-bar}

\begin{figure*}
\centering
	\includegraphics[width=15cm]{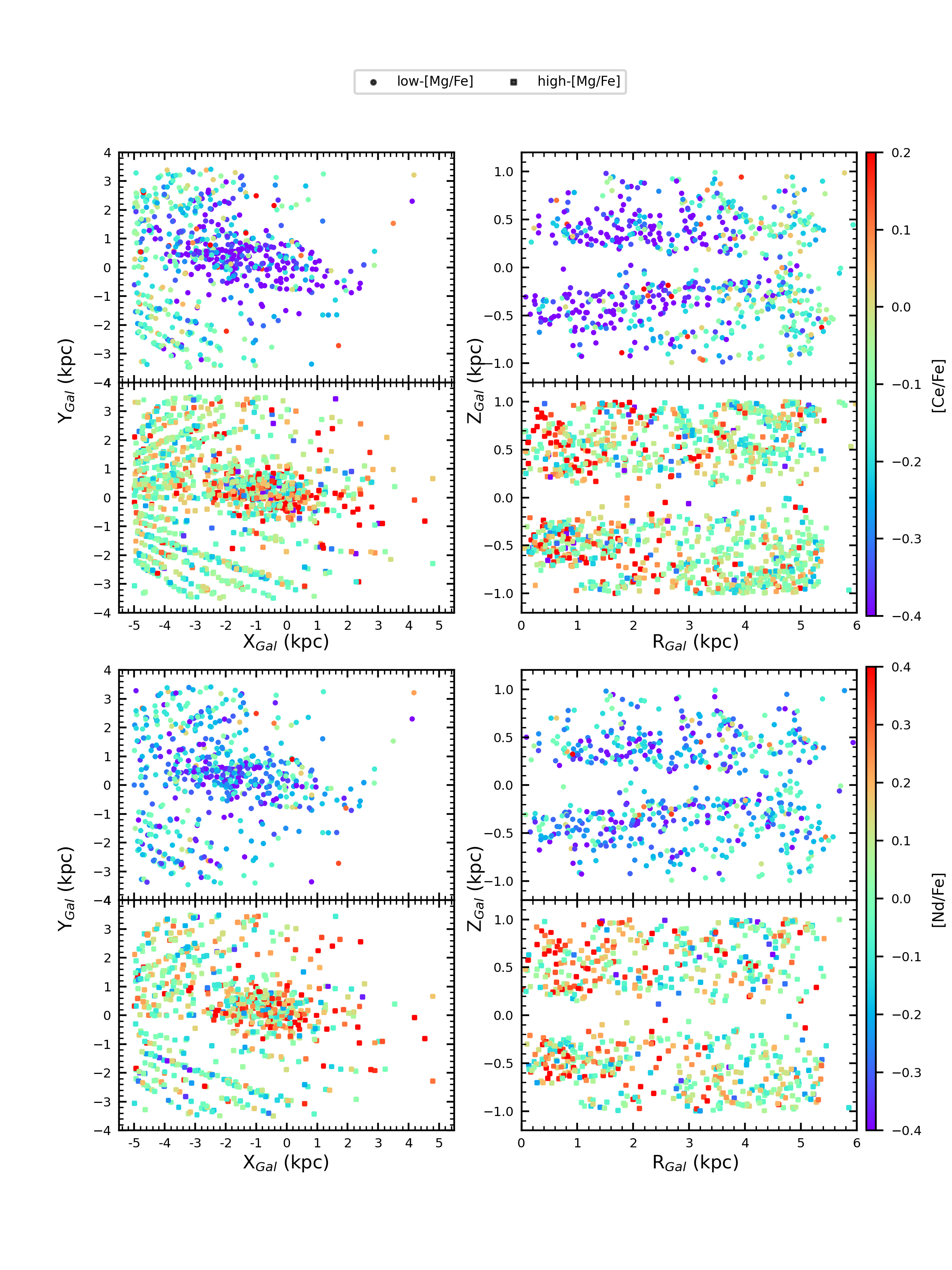}
    \caption{Distribution of bulge-bar-BAWLAS stars in the X$_{Gal}$-Y$_{Gal}$ and R$_{Gal}$-Z$_{Gal}$ planes, color-coded according to their [Ce/Fe] and [Nd/Fe] ratios. Circles represent the low-[Mg/Fe] population; Squares represent the high-[Mg/Fe] population. 
    }
    \label{X_vs_Y_color_CeFe}
\end{figure*}

\begin{figure*}
\centering
	\includegraphics[width=15cm]{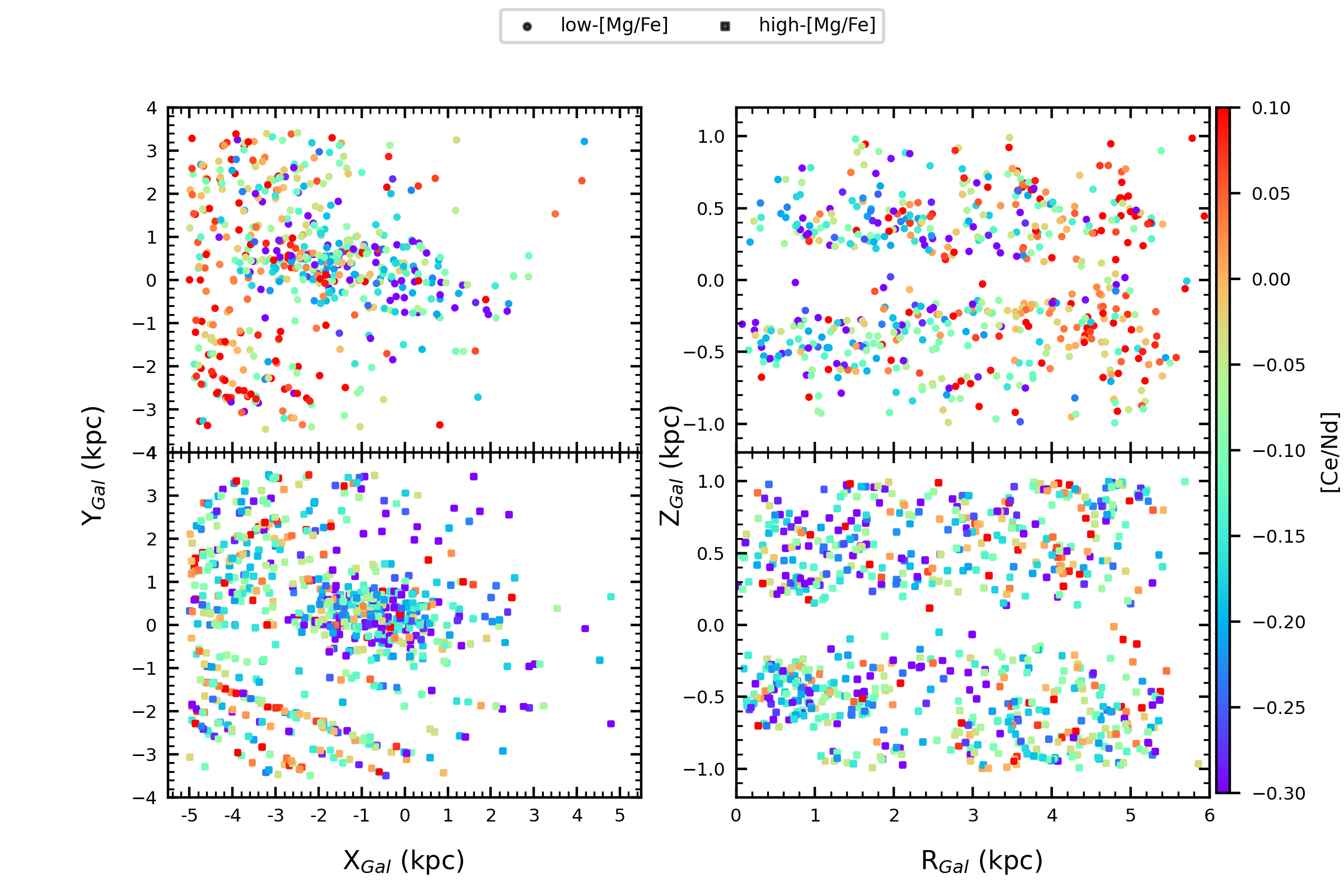}
    \caption{Distribution of bulge-bar-BAWLAS stars in the X$_{Gal}$-Y$_{Gal}$ and R$_{Gal}$-Z$_{Gal}$ planes, color-coded according to their [Ce/Nd] ratios. The top panels represent the low-[Mg/Fe] population (circles); the bottom panels represent the high-[Mg/Fe] population (squares). 
    }
    \label{X_vs_Y_color_CeNd}
\end{figure*}

\begin{figure*}
\centering
	\includegraphics[width=15cm]{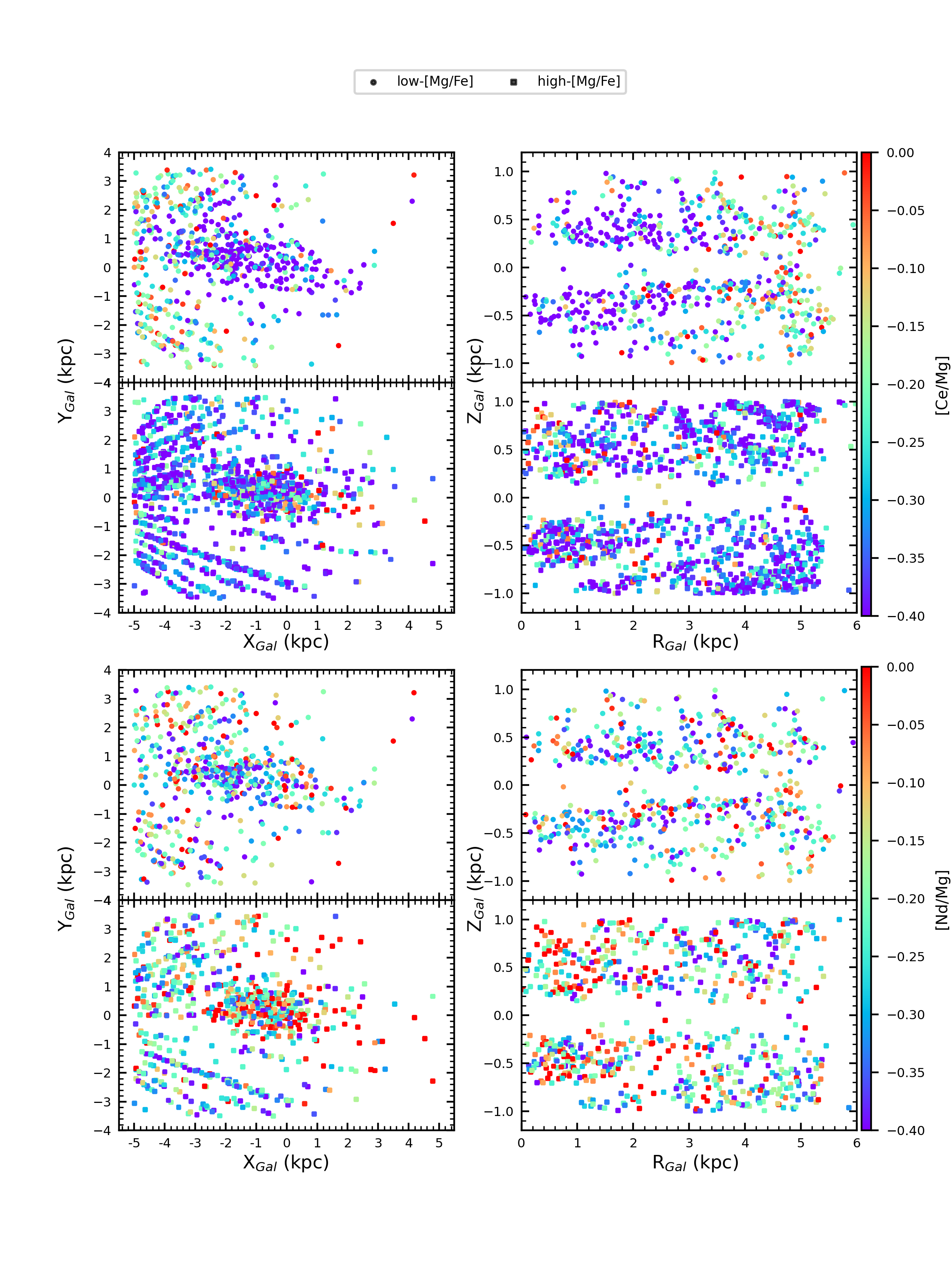}
    \caption{Distribution of bulge-bar-BAWLAS stars in the X$_{Gal}$-Y$_{Gal}$ and R$_{Gal}$-Z$_{Gal}$ planes, color-coded according to their [Ce/Mg] and [Nd/Mg] ratios. Circles represent the low-[Mg/Fe] population; Squares represent the high-[Mg/Fe] population. 
    }
    \label{X_vs_Y_color_CeMg}
\end{figure*}

The spatial distribution of stars in the bulge-bar can add to our understanding of their formation and evolution. The X-shaped structure of the bulge \citep{McWilliam2010,Wegg2013,Ness2016} 
probably indicates a secular bulge formation with a long period of quiet MW evolution without major mergers that could disturb the orbits of the bulge stars. In particular, the metal-rich (0.0$\leq$[Fe/H]$\leq$0.5) bulge population shows clearly this X-shaped structure \citep{Portail2017} and a boxy distribution \citep{Zoccali2017}. On the other hand, the metal-poor bulge population is more centrally concentrated \citep{Zoccali2017}, with an X-shape that is less evident \citep{Portail2017}, and it is dominated by an $\alpha$-rich population \citep{Queiroz2021}. 

To answer the question of how the chemical patterns of the neutron capture elements are distributed in the inner galaxy, we map the Ce and Nd abundances of the bulge-bar-BAWLAS sample using distances and Galactic coordinates (X$_{Gal}$, Y$_{Gal}$, and Z$_{Gal}$) provided by \citet{Queiroz2021}. They estimated the star positions and distances through the Bayesian StarHorse code \citep{Santiago2016,Queiroz2018}  using APOGEE and Gaia EDR3 data with typical distance uncertainties of around 7\%. The bulge-bar-BAWLAS sample is spatially contained within the region of $|X_{Gal}|<$5 kpc, $|Y_{Gal}|<$3.5 kpc, and $|Z_{Gal}|<$1 kpc, with the Galactocentric distances (R$_{Gal}$) ranging between 0.0 and 6.0 kpc.

\subsubsection{X$_{Gal}$-Y$_{Gal}$ and Z$_{Gal}$-R$_{Gal}$ projections}

In Figures \ref{X_vs_Y_color_CeFe}, and \ref{X_vs_Y_color_CeNd}, we show the X$_{Gal}$-Y$_{Gal}$ (face-on) and Z$_{Gal}$-R$_{Gal}$ (edge-on) projections color-coded by [Ce/Fe], [Nd/Fe] and [Ce/Nd] ratios, respectively, for the low-(circles) and high-[Mg/Fe] (squares) populations of the bulge-bar-BAWLAS sample. The low-[Mg/Fe] population is dominated by very low [Ce/Fe] and [Nd/Fe] values (dark blue circles) in the region close to the center of the Milky Way. This panorama begins to change beyond R$_{Gal}\approx$3.5 kpc, with the low-[Mg/Fe] stars, in general, showing higher [Ce/Fe] and [Nd/Fe] ratios and being dominated by stars with $-$0.2$\lesssim$[Ce/Fe]$\lesssim$0.0 and $-$0.1$\lesssim$[Nd/Fe]$\lesssim$0.1 (light blue or green circles in Figure \ref{X_vs_Y_color_CeFe}). Looking at the distribution of the [Ce/Nd] ratio in the low-[Mg/Fe] population (top panels of Figure \ref{X_vs_Y_color_CeNd}), the central region is dominated by low [Ce/Nd] values (blue and green circles), with the outermost region being occupied mostly by stars with [Ce/Nd]$\geq$0 (red circles). These maps indicate that the Ce and Nd production in the bulge-bar central region is dominated by the r-process, while in regions further away from the center the s-process contribution dominates.

The high-[Mg/Fe] population displays a distinct picture from the one discussed above for the low-[Mg/Fe] stars. Squares in Figure \ref{X_vs_Y_color_CeFe} show that the high-[Mg/Fe] stars with high [Ce/Fe] and [Nd/Fe] values (red squares) dominate the most central region of the MW. While those stars with $-$0.2$\lesssim$[Ce/Fe]$\lesssim$0.05 and $-$0.1$\lesssim$[Nd/Fe]$\lesssim$0.15 (orange and green squares) are dominant beyond Galactocentric distances of R$_{Gal}$$\sim$3 kpc.
For [Ce/Nd] ratios in the high-[Mg/Fe] population (bottom panels of Figure \ref{X_vs_Y_color_CeNd}), stars with the low [Ce/Nd] values (blue and green squares) dominate the most central region of the MW, which is a similar pattern as for the low-[Mg/Fe] stars. These results suggest a greater efficiency in the Nd enrichment compared to Ce in this region, which is a signature of the r-process.

We also map the [Ce/Mg] and [Nd/Mg] ratios in the bulge-bar (Figure \ref{X_vs_Y_color_CeMg}) to investigate correlations between the studied neutron capture elements and the pristine core-collapse supernovae element, Mg. 
Stars with very low [Ce/Mg] and [Nd/Mg] ratios (blue circles) occupy the central regions out to about R$_{Gal}$$\sim$4 kpc for the low-[Mg/Fe] population (circles in Figure \ref{X_vs_Y_color_CeMg}). 
On the other hand, for the high-[Mg/Fe] population (squares in Figure \ref{X_vs_Y_color_CeMg}), the stars with very low [Ce/Mg] values (blue squares) are more dispersed throughout the inner Galactic region. The high-[Mg/Fe] population also has a significant number of stars with [Nd/Mg]$\approx$0.0 (red squares) in the most central region of the bulge. 

Summarizing, Figures \ref{X_vs_Y_color_CeFe} - \ref{X_vs_Y_color_CeMg} indicate spatial dependencies of the neutron-capture elemental abundance ratios relative to Fe and Mg in the bulge-bar, as found by \citet{Queiroz2021} for alpha elements and metallicities. In addition, we detected that such dependencies vary with the bulge population.

\section{Conclusions}

The current era of large spectroscopic surveys coupled with Gaia data transformed our view of the Galaxy. The inner Galaxy, including the bulge-bar, was specifically targeted by the APOGEE survey using high-resolution near-infrared spectroscopy to reveal its chemistry and structure details. 
The discovery of absorption lines of the two heavy elements Cerium and Neodymium in the H-Band spectra of APOGEE opened a new window in the chemical tagging provided by APOGEE, allowing the study of the chemical evolution of the s- and r-process in the MW. However, these lines tend to be weak and blended with other lines, requiring careful analysis. Here, we used the Ce and Nd abundances from the BACCHUS Analysis of Weak Lines \citep[BAWLAS][]{Hayes2022} APOGEE DR 17 catalog to investigate the chemical pattern of these elements in the Galactic bulge-bar.

Our sample consists of approximately 2000 bulge-bar stars obtained through cross-matching the BAWLAS catalog with the sample from \citet{Queiroz2021}, which selected bulge stars cleaned of foreground (disc or halo) stars using the reduced-proper-motion (RPM) diagram. Before analyzing the neutron-capture elements, we chemically characterized our sample through their Fe, Mg, Al, and N abundances, classifying the stars as low- or high-[Mg/Fe], or most metal-poor ([Fe/H]$<-$1.2), or chemically peculiar stars. Also, we performed cross-matches between our sample and some catalogs of known bulge populations 
(\citep[N-rich stars,][]{Fernandez-Trincado2022}; 
\citep[GC stars,][]{VasilievBaumgardt2021}; 
\citep[the Heracles substructure,][]{Horta2021}) 
to identify other stellar populations present in our sample.  This comparison identified 14 N-rich stars, 6 probable GC member stars (one from Liller 1, one from NGC 6380, two from NGC 6522, one from NGC 6569, and one from Pismis 26), and 18 Heracles stars. This separation of our sample into different populations allowed a more accurate interpretation of the chemical pattern of the neutron-capture elements.

Within the bulge-bar-BAWLAS stars, we found a variation in [Ce/Fe] as a function of [Fe/H] for the low- and high-[Mg/Fe] populations that reveal a complex relationship between the [Ce/Fe] ratio and [Fe/H].  In particular, the high-[MgFe] population displays a ``knee'' feature at [Fe/H]$\approx-$1.0.

In the [Nd/Fe]-[Fe/H] plane, the low- and high-[Mg/Fe] populations exhibit a simple linear trend, with the [Nd/Fe] ratio decreasing as the metallicity increases. This relationship between the [Nd/Fe] ratio and [Fe/H] is similar to that obtained for Eu, an r-process dominated element, in the Galactic bulge \citep{Johnson2012,VanderSwaelmen2016}. Corroborating this scenario, [Ce/Nd]$<$0 for most bulge-bar stars (mainly for the high-[Mg/Fe] population) indicates a significant contribution from the r-process, greater than in the Sun, for Ce and Nd.

We also investigated the Ce and Nd abundances relative to Mg (a dominant core-collapse supernova element) and Mn (a dominant type Ia supernova element) as functions of [Fe/H].  In general, the [Ce/X] and [Nd/X] ratios for Mg and Mn also show an increase as the metallicity decreases. For the high-[Mg/Fe] population, we found a change in the trend between the [Ce/Mg] ratio and [Fe/H] around [Fe/H]$\approx-$0.5, probably caused by late Ce enrichment in the interstellar medium by low-mass AGB stars.

The most metal-poor stars ([Fe/H]$<-$1.2) present [Ce/Fe]$>$0.0, [Ce/Mn]$>$0.0, and [Ce/Mg]$<$0.0. Only one of these stars had a measured Nd abundance, with [Nd/Fe]=+0.47 and [Nd/Mg]=+0.16. Overall, GC, Heracles, CP, and N-rich stars in our sample overlap with the high-[Mg/Fe] population in the [Ce/Fe]-[Fe/H] and [Nd/Fe]-[Fe/H] planes. From Ce and Nd's point of view, Heracles stars in our sample do not present a chemical pattern distinct from other bulge-bar stars with similar metallicities.

Through the X$_{Gal}$-Y$_{Gal}$ and Z$_{Gal}$-R$_{Gal}$ Galaxy projections, we found a spatial chemical dependence of the Ce and Nd abundances for our sample, with low- and high-[Mg/Fe] populations showing distinct dependencies. 
In the region close to the center of the MW, the low-[Mg/Fe] population is dominated by stars with low [Ce/Fe], [Ce/Mg], [Nd/Mg], [Nd/Fe], and [Ce/Nd] ratios. The low [Ce/Nd] ratio indicates a significant contribution in this central region from r-process yields for the low-[Mg/Fe] population, unlike the region further from the center, which is dominated by stars with high [Ce/Nd] ratios (s-process yields). On the other hand, the low [Ce/X], and [Nd/X] ratios relative to Fe and Mg suggest a greater contribution from core-collapse and type Ia supernovae yields compared to the r-process astrophysical site yields in the chemical pattern of the low-[Mg/Fe] population located in the central bulge-bar region.

The high-[Mg/Fe] population displays a scenario opposite to that shown by the low-[Mg/Fe] stars for the [Ce/Fe], and [Nd/Fe] ratios, with the most central region of the bulge-bar dominated by stars with high [Ce/Fe] and [Nd/Fe] ratios. However, in this Galactic region, the [Ce/Nd] ratios present lower values for the high-[Mg/Fe] population, as also found by low-[Mg/Fe] stars. The high-[Mg/Fe] population also shows a significant number of stars with [Nd/Mg] ratios close to solar values in the central region of the bulge-bar. 

Our results highlight the complexity of chemical evolution in the bulge-bar, which, in addition to being embedded in multiple stellar populations with different histories, reveal intriguing relationships between the chemical abundances of neutron-capture elements with both the metallicity and the positions of the stars. This interesting spatial distribution dependence is shared by other elements, such as iron and alpha elements \citep{Queiroz2021}. The multiple nucleosynthesis pathways for producing neutron-capture elements challenge our understanding of the astrophysical sources of these elements in the bulge-bar populations. The low [Ce/Nd] ratios present in the low metallicity bulge-bar stars indicate an ancient enrichment dominated by the r-process, possibly a signature of magneto-rotational supernovae yields generated through massive stars. This r-process astrophysical site occurs on a short enough timescale to enrich these stars, as pointed out by \citet{Lucey2022}. Our results show a greater efficiency of this enrichment in the most central region of the bulge. Beyond Ce and Nd, the chemical abundances of other heavy elements in bulge-bar stars also indicate a significant contribution of the r-process at low metallicity \citep{VanderSwaelmen2016,Duong2019,Lucey2022}. Some studies also point to an important contribution of spinstars (massive rotating stars, an s-process site) in the abundance of neutron-capture elements of metal-poor bulge stars \citep{Chiappini2011,Prantzos2018,Razera2022}.

Over time, the contribution of s-process yields becomes more significant in Bulge stars since the main source of these products, low-mass AGB stars, presents Gigayear time delays in the enrichment of the interstellar medium. The [Ce/Nd]$>$0.0 for some metal-rich stars in the low-[Mg/Fe] population points to this late s-process enrichment in neutron-capture elements.  It is important to note, however, that these stars are located in the outermost regions of the inner Galaxy, as shown in the left top panel of Figure \ref{X_vs_Y_color_CeNd}.


\acknowledgments

We thank the referee for suggestions that improved the paper. JVSS acknowledges the PDS/FAPERJ programme under grant E-26/203.899/2022. SD acknowledges CNPq/MCTI for grant 306859/2022-0. DS thanks the National Council for Scientific and Technological Development – CNPq. RG acknowledges PDR-10/FAPERJ grant E26-205.964/2022. TM acknowledges financial support from the Spanish Ministry of Science and Innovation (MICINN) through the Spanish State Research Agency, under the Severo Ochoa Program 2020-2023 (CEX2019-000920-S).
JGF-T gratefully acknowledges the grants support provided by Agencia Nacional de Investigaci\'on y Desarrollo de Chile (ANID) Fondecyt Iniciaci\'on No. 11220340, ANID Fondecyt Postdoc No. 3230001 (Sponsoring researcher), from two Joint Committee ESO-Government of Chile grants under the agreements 2021 ORP 023/2021 and 2023 ORP 062/2023, and from Becas Santander Movilidad Internacional Profesores 2022, Banco Santander Chile.
MZ is supported by ANID, Millenium Science Initiative, ICN12\_009 awarded to the Millennium Institute of Astrophysics (M.A.S.), by the ANID BASAL Center for Astrophysics and Associated Technologies (CATA) through grant FB210003, and by FONDECYT Regular grant No. 1230731

Funding for the Sloan Digital Sky Survey IV has been provided by the Alfred P. Sloan Foundation, the U.S. Department of Energy Office of Science, and the Participating Institutions. SDSS-IV acknowledges support and resources from the Center for High-Performance Computing at the University of Utah. The SDSS website is www.sdss.org.

SDSS-IV is managed by the Astrophysical Research consortium for the Participating Institutions of the SDSS Collaboration including the Brazilian Participation Group, the Carnegie Institution for Science, Carnegie Mellon University, the Chilean Participation Group, the French Participation Group, HarvardSmithsonian Center for Astrophysics, Instituto de Astrofísica de Canarias, The Johns Hopkins University, Kavli Institute for the Physics and Mathematics of the Universe (IPMU)/University of Tokyo, Lawrence Berkeley National Laboratory, Leibniz Institut fur Astrophysik Potsdam (AIP), Max-Planck-Institut fur Astronomie (MPIA Heidelberg), Max-Planck Institut fur Astrophysik (MPA Garching), Max-Planck-Institut fur Extraterrestrische Physik (MPE), National Astronomical Observatory of China, New Mexico State University, New York University, University of Notre Dame, Observatório Nacional/MCTI, The Ohio State University, Pennsylvania State University, Shanghai Astronomical Observatory, United Kingdom Participation Group, Universidad Nacional Autónoma de México, University of Arizona, University of Colorado Boulder, University of Oxford, University of Portsmouth, University of Utah, University of Virginia, University of Washington, University of Wisconsin, Vanderbilt University, and Yale University.

This research made use of following Python packages: {\tt matplotlib} \citep[][]{Hunter2007}, {\tt Numpy} \citep[][]{Harris2020}, {\tt Scipy} \citep[][]{Virtanen2020}.

%

\vspace{5mm}
\facilities{Sloan (APOGEE)}


\software{\tt matplotlib} \citep[][]{Hunter2007}, {\tt Numpy} \citep[][]{Harris2020}, {\tt Scipy} \citep[][]{Virtanen2020}

\bibliography{reference}




\end{document}